\documentclass[twocolumn,aps,prc,superscriptaddress,showpacs,floatfix,longbibliography,nofootinbib]{revtex4-1}
\usepackage{url}
\usepackage{cancel}
\usepackage[colorlinks,linkcolor=blue,citecolor=blue,filecolor=black,urlcolor=blue]{hyperref}
\usepackage{epsfig,graphics}
\usepackage{graphicx}
\usepackage{dcolumn}
\usepackage{bm}
\usepackage[usenames]{color}
\usepackage{amssymb}
\usepackage{amsmath}
\usepackage{multirow}
\usepackage{float}
\usepackage{harpoon}
\usepackage{MnSymbol}
\usepackage{appendix}
\usepackage{color}
\usepackage{hyperref}
\usepackage{cleveref}
\usepackage{dutchcal}

\begin{document}

\title{Spin dynamics in intermediate-energy heavy-ion collisions with rigorous angular momentum conservation
}
\author{Rong-Jun Liu}
\affiliation{Shanghai Institute of Applied Physics, Chinese Academy of Sciences, Shanghai 201800, China}
\affiliation{University of Chinese Academy of Sciences, Beijing 100049, China}
\author{Jun Xu}\email[Correspond to\ ]{junxu@tongji.edu.cn}
\affiliation{School of Physics Science and Engineering, Tongji University, Shanghai 200092, China}
\affiliation{Shanghai Institute of Applied Physics, Chinese Academy of Sciences, Shanghai 201800, China}
\begin{abstract}
We have revisited the spin dynamics in intermediate-energy heavy-ion collisions based on the improved spin- and isospin-dependent Boltzmann-Uehling-Uhlenbeck transport model, particularly with the constraint of rigorous angular momentum conservation incorporated. We have studied the spin polarization of free nucleons and tritons/$^3$He as well as the spin alignment of deuterons, and predicted the flow splittings for their different spin states. We have also demonstrated that the spin-dependent potential may enhance dissipations and thus have a non-negligible effect on the spin-averaged transverse flow at low collision energies. When rigorous angular momentum conservation in each spin-dependent nucleon-nucleon collision is incorporated, it affects the overall dynamics, the flow, and also the spin polarization, while the effects of the spin-orbit potential on the spin-related observables are still appreciable. The well-developed SIBUU model could be further extended to include hyperons or vector mesons, or used as a hadronic afterburner for spin-related studies in relativistic heavy-ion collisions, with more inelastic channels incorporated in the future.
\end{abstract}
\maketitle

\section{Introduction}
\label{introduction}

Spin dynamics is of general interest in various fields. In the nuclear physics community, the spin and chiral dynamics in relativistic heavy-ion collisions have been under extensive investigations in the past fifteen years~\cite{Kharzeev:2015znc,Huang:2015oca,Becattini:2020ngo}. Due to the high energy density reached in relativistic heavy-ion collisions, the chiral symmetry is restored for both quarks and antiquarks, and they can be considered as massless particles, whose spins are coupled with their momenta depending on their helicities. Under the strong magnetic field and vorticity field produced in relativistic heavy-ion collisions, the dynamics of these partons exhibits various chiral anomalies, which have been proposed theoretically~\cite{Fukushima:2008xe,Burnier:2011bf,Kharzeev:2007tn} and investigated experimentally~\cite{STAR:2009wot,ALICE:2012nhw,STAR:2015wza,Zhao:2014aja}. More recently, the spin polarizations of $\Lambda$ and $\bar\Lambda$ hyperons~\cite{STAR:2022fan} and the spin alignment of vector mesons~\cite{STAR:2022fan} have been measured through the angular distribution of their decays, and these have attacked considerable attention and led to various related studies, e.g., effects of the hadronic evolution on the spin observables~\cite{Becattini:2019ntv,Xia:2019fjf}.

Compared to the extensive investigations in high-energy nuclear physics community, the spin dynamics in low- and intermediate-energy heavy-ion collisions have not attracted much attention. Since nucleons are massive particles, their spins and momenta are decoupled, and their spin dynamics is mostly dominated by the nuclear spin-orbit interaction and to some extent affected by the nuclear tensor force (see, e.g., Ref.~\cite{Xu:2015kxa} for a review). Both the spin-orbit interaction and the tensor force are under hot investigations in nuclear structure studies and may largely affect the shell evolution of finite nuclei~\cite{Mayer:1948zz,Mayer:1949pd,Haxel:1949fjd,Otsuka:2005zz,Otsuka:2006zz,Otsuka:2009qs}. In low-energy reactions, the spin-orbit nuclear interaction may lead to the internal spin excitation of the colliding nuclei, enhance the dissipation, increase the fusion threshold, and affect significantly the fusion cross section based on the time-dependent Hartree-Fock approach~\cite{PhysRevLett.56.2793,Maruhn:2006uh,Reinhard:1988zz}, while the tensor force may have subdominant effects~\cite{Stevenson:2015dva,Godbey:2019vlg}. In intermediate-energy heavy-ion collisions, we have developed a spin- and isospin-dependent Boltzmann-Uehling-Uhlenbeck (SIBUU) transport model~\cite{Xu:2012hh,Xia:2016xiw} by incorporating spin degree of freedom into the IBUU transport model. We have observed different collective flows for nucleons of different spin states with respect to $y$ direction perpendicular to the reaction plane~\cite{Xu:2012hh,Xia:2014qva}. We have also observed global and local spin polarizations~\cite{Xia:2019whr} in $y$ direction and along the beam direction, respectively. These phenomena originate from different forces on nucleons of different spins, a general feature called the spin-Hall effect~\cite{PhysRevLett.83.1834} for particle transport under the spin-orbit interaction.

In the present study, we have further improved the SIBUU transport model by incorporating the constraint of rigorous angular momentum conservation. We have also made minor improvements on the treatment of spin-dependent nucleon-nucleon collisions and the spin-dependent coalescence procedure for the production of light clusters with different spins. While the spin dynamics remains qualitatively similar compared to our previous studies, we pay more attention to some new features in the present work. First of all, we find that the spin polarization as well as the splitting of the collective flows for free nucleons can be affected by the constraint of angular momentum conservation. Second, besides the spin polarization of nucleons and tritons/$^3$He with spin $\frac{1}{2}$, the spin alignment of deuterons with spin $1$ can also be observed. Third, we find that the spin-orbit potential has a non-negligible effect on the spin-averaged transverse flow at low collision energies.

The rest part of the paper is organized as follows. We describe the details of the SIBUU transport model as well as the improvements, especially the incorporation of the angular momentum conservation, in Sec.~\ref{theory}. We investigate the spin polarization or spin alignment of free nucleons and light clusters, as well as the splitting of collective flows for their different spin states, and compare the spin-averaged transverse flow with and without spin-orbit potential, in Sec.~\ref{results}. We conclude and give an outlook in Sec.~\ref{summary}.

\section{Theoretical Framework}
\label{theory}

In this section, we first review the basic framework of the SIBUU transport model. Next, we describe in detail how we incorporate the constraint of rigorous angular momentum conservation in each nucleon-nucleon collision and in the mean-field evolution. We also present the improved treatments for spin-dependent nucleon-nucleon collisions and coalescence procedures.

\subsection{Basic framework of SIBUU}

The SIBUU transport model solves the spin-dependent BUU equation written as~\cite{Ring1980,Smith1989}
\begin{eqnarray}\label{BLE}
\frac{\partial \hat{f}}{\partial t}&+&\frac{i}{\hbar}\left [ \hat{\varepsilon},\hat{f}\right]+\frac{1}{2}\left ( \frac{\partial \hat{\varepsilon}}{\partial \vec{p}}\cdot \frac{\partial \hat{f}}{\partial \vec{r}}+\frac{\partial \hat{f}}{\partial \vec{r}}\cdot \frac{\partial \hat{\varepsilon}}{\partial \vec{p}}\right ) \nonumber\\
&-&\frac{1}{2}\left ( \frac{\partial \hat{\varepsilon}}{\partial
\vec{r}}\cdot \frac{\partial \hat{f}}{\partial
\vec{p}}+\frac{\partial \hat{f}}{\partial \vec{p}}\cdot
\frac{\partial \hat{\varepsilon}}{\partial \vec{r}}\right )=I_c,
\end{eqnarray}
where the single-particle energy $\hat{\varepsilon}$ and the phase-space distribution $\hat{f}$ are $2\times 2$ matrics
\begin{eqnarray}
\hat{\varepsilon}(\vec{r},\vec{p})&=&\varepsilon(\vec{r},\vec{p})\hat{I}+\vec{h}(\vec{r},\vec{p})\cdot \vec{\sigma},\label{ener} \\
\hat{f}(\vec{r},\vec{p}) &=&
f_{0}(\vec{r},\vec{p})\hat{I}+\vec{g}(\vec{r},\vec{p})\cdot\vec{\sigma},
\label{dens}
\end{eqnarray}
with $\vec{\sigma}=(\sigma_{x},\sigma_{y},\sigma_{z})$ and $\hat{I}$ being
respectively the Pauli matrices and the $2\times2$ unit matrix. In the above, $\varepsilon(\vec{r},\vec{p})$ and $f_{0}(\vec{r},\vec{p})$ are the spin-averaged single-particle energy and phase-space distribution, and $\vec{h}(\vec{r},\vec{p})$ and $\vec{g}(\vec{r},\vec{p})$ represent their spin-dependent contributions, respectively. $I_c$ represents the collision term.

We start from the Skyrme-type spin-orbit nuclear interaction between nucleons at $\vec{r}_1$ and $\vec{r}_2$ expressed as~\cite{Vautherin:1971aw}
\begin{equation}
v_{so} = i W_0 (\vec{\sigma}_1+\vec{\sigma}_2) \cdot \vec{k}^\prime \times
\delta(\vec{r}_1-\vec{r}_2) \vec{k},
\end{equation}
where $W_0$ is the strength of the spin-orbit coupling whose default value is set to be 133 MeVfm$^5$~\cite{PhysRevC.82.024321} in this study, $\vec{\sigma}_{1(2)}$ is the Pauli matrices, $\vec{k}=(\vec{p}_1-\vec{p}_2)/2$ is the relative momentum operator acting on the right with $\vec{p}=-i\nabla$, and $\vec{k}^\prime$ is the complex conjugate of $\vec{k}$. Based on the Hartree-Fock method~\cite{Engel:1975zz}, the contribution to the energy-density functional can be expressed as
\begin{eqnarray}\label{vso}
V_{so} &=& V_{so}^0 + \sum_\tau V_{so}^\tau \notag\\
&=&-\frac{W_0}{2}[\rho \nabla \cdot \vec{J} + \vec{s} \cdot \nabla \times \vec{j} \notag\\
&+& \sum_\tau (\rho_\tau \nabla \cdot \vec{J}_\tau + \vec{s}_\tau \cdot \nabla \times \vec{j}_\tau)],
\end{eqnarray}
where both time-even and time-odd contributions are included. From the variational principle and in the semi-classical approximation, the contribution of the spin-orbit interaction to the single-particle energy can be written as
\begin{equation}
\hat{h}_{so} = \varepsilon_{so} + \vec{h} \cdot \vec{\sigma},
\end{equation}
with
\begin{eqnarray}
\varepsilon_{so} &=& -\frac{W_{0}}{2}\nabla\cdot (\vec{J}+\vec{J}_{\tau}) -\frac{W_{0}}{2}[\vec{p} \cdot (\nabla\times (\vec{s}+\vec{s}_{\tau}))], \\
\vec{h} &=& -\frac{W_{0}}{2}\nabla \times (\vec{j}+\vec{j}_{\tau}) +\frac{W_{0}}{2}[\nabla(\rho+\rho_{\tau}) \times \vec{p}].
 \label{hso}
\end{eqnarray}
In the above, $\tau=n,p$ is the isospin index, $\rho=\sum_\tau \rho_\tau$, $\vec{s}=\sum_\tau \vec{s}_\tau$, $\vec{j}=\sum_\tau \vec{j}_\tau$, and $\vec{J}=\sum_\tau \vec{J}_\tau$ are the number, spin, momentum, and spin-current densities, respectively. Omitting the isospin index, these densities can be expressed in terms of $f_0$ and $\vec{g}$ in Eq.~(\ref{dens}) as~\cite{Xia:2016xiw}
\begin{eqnarray}
\rho(\vec{r}) &=& 2 \int d^{3}p f_0(\vec{r},\vec{p}), \label{rhor}\\
\vec{s}(\vec{r}) &=& 2 \int d^{3}p \vec{g}(\vec{r},\vec{p}),  \\
\vec{j}(\vec{r}) &=& 2 \int d^{3}p \frac{\vec{p}}{\hbar}f_0(\vec{r},\vec{p}),  \\
\vec{J}(\vec{r}) &=& 2 \int d^{3}p \frac{\vec{p}}{\hbar} \times
\vec{g}(\vec{r},\vec{p}). \label{Jr}
\end{eqnarray}
The spin-averaged energy density in Eq.~(\ref{ener}) is
\begin{equation}
\varepsilon = \frac{p^2}{2m} + U_{MID} + \varepsilon_{so}
\end{equation}
with
\begin{equation}
U_{MID} = a\left(\frac{\rho}{\rho_0}\right)+b\left(\frac{\rho}{\rho_0}\right)^c \pm 2E_{sym}^{pot}\left(\frac{\rho}{\rho_0}\right)^{\gamma_{sym}} \left(\frac{\rho_n-\rho_p}{\rho}\right)
\end{equation}
being the momentum-independent mean-field potential, with the ``$+(-)$" sign for neutrons (protons), and parameters $a=-209.2$ MeV, $b=156.4$ MeV, $c=1.35$, $E_{sym}^{pot}=18$ MeV, and $\gamma_{sym}=2/3$ that reproduce the empirical nuclear matter properties.

As shown in Ref.~\cite{Xia:2016xiw}, the left-hand side of the spin-dependent BUU equation can be solved if the time evolution of the $i$th test particle is simulated according to
\begin{eqnarray}
\frac{d\vec{r}_i}{dt} &=& \frac{\partial }{\partial \vec{p}_i} (\varepsilon + \vec{h} \cdot \vec{\sigma}_i),  \label{eom1}\\
\frac{d\vec{p}_i}{dt} &=& - \frac{\partial }{\partial \vec{r}_i} (\varepsilon + \vec{h} \cdot \vec{\sigma}_i), \label{eom2}\\
\frac{d\vec{\sigma}_i}{dt} &=& 2 \vec{h} \times \vec{\sigma}_i. \label{eom3}
\end{eqnarray}
In the above, $\vec{\sigma}_i$ is now a unit vector describing the expectation direction of the nucleon spin in the semiclassical case, and the third equation describes the precession of $\vec{\sigma}_i$ in the spin-dependent mean-field potential. Different forces on nucleons of different spins can be seen from Eq.~(\ref{eom2}). In the present study, we extend the above framework by using the lattice Hamiltonian method~\cite{Lenk:1989zz}, so that the average number, spin, momentum, and spin-current densities at the sites of a three-dimensional cubic lattice are calculated respectively as~\cite{Xia:2019whr}
\begin{eqnarray}
\rho_L(\vec{r}_{\alpha})&=&\sum_{i}S(\vec{r}_{\alpha}-\vec{r}_i),\\
\vec{s}_L(\vec{r}_{\alpha})&=&\sum_{i}\vec{\sigma}_iS(\vec{r}_{\alpha}-\vec{r}_i),\\
\vec{j}_L(\vec{r}_{\alpha})&=&\sum_{i}\vec{p}_iS(\vec{r}_{\alpha}-\vec{r}_i),\\
\vec{J}_L(\vec{r}_{\alpha})&=&\sum_{i}\left(\vec{p}_i \times \vec{\sigma}_i\right)S(\vec{r}_{\alpha}-\vec{r}_i).
\end{eqnarray}
In the above, $\alpha$ is a site index and $\vec{r}_{\alpha}$ is the position of site $\alpha$, and $S$ is the shape function describing the contribution of a test particle at $\vec{r}_i$ to the average density at $\vec{r}_{\alpha}$, i.e.,
\begin{eqnarray}
S(\vec{r})=\frac{1}{N(nl)^6}g(x)g(y)g(z)
\end{eqnarray}
with
\begin{eqnarray}
g(q)=(nl-|q|)\Theta(nl-|q|).
\end{eqnarray}
$N$ is the number of parallel events, $l$ is the lattice spacing, $n$ determines the range of $S$, and $\Theta$ is the Heaviside function. We adopt $N=400$, $l=1$ fm, and $n=2$ in the present study. The Hamiltonian of the whole system is
\begin{equation}
H=\sum_{i}\frac{\vec{p}_{i}^{2}}{2m}+Nl^3\sum_\alpha [V_{MID}(\vec{r}_{\alpha}) + V_{so}(\vec{r}_{\alpha})]
\end{equation}
where $V_{MID}$ is the corresponding potential energy density from $U_{MID}$, and $V_{so}$ is the potential energy density from the spin-orbit interaction [Eq.~(\ref{vso})]. The canonical equations of motion in the lattice Hamiltonian framework now become
\begin{eqnarray}
\frac{d\vec{r}_i}{dt} &=& \frac{\partial H}{\partial \vec{p}_i}, \\
\frac{d\vec{p}_i}{dt} &=& - \frac{\partial H}{\partial \vec{r}_i},\\
\frac{d\vec{\sigma}_i}{dt} &=& \frac{1}{i} [\vec{\sigma}_i, H], \label{proc}
\end{eqnarray}
which are similar to Eqs.~(\ref{eom1}-\ref{eom3}), but with a smearing of the test-particle size and more accurate numerical derivatives.

\subsection{Spin-dependent nucleon-nucleon collisions}

The above part describes the spin-dependent mean-field evolution. For the collision term $I_c$ in Eq.~(\ref{BLE}), we also treat it as spin-dependent in SIBUU. We only consider elastic nucleon-nucleon collisions, and all inelastic collisions are turned off in the present study.

We use different spin-singlet and spin-triplet neutron-neutron (proton-proton) and neutron-proton elastic scattering cross sections as parameterized in Ref.~\cite{Xia:2017dbx}, which were extracted from the phase-shift analyses of nucleon-nucleon scatterings in free space~\cite{PhysRevC.15.1002}. For two nucleons with their expectation spin directions at
\begin{eqnarray}
\vec{\sigma}_1=(\sin\theta_1\cos\phi_1,\sin\theta_1\sin\phi_1,\cos\theta_1), \notag\\
\vec{\sigma}_2=(\sin\theta_2\cos\phi_2,\sin\theta_2\sin\phi_2,\cos\theta_2),
\end{eqnarray}
where $\theta_{1(2)}$ and $\phi_{1(2)}$ are the polar angle and azimuthal angle, respectively, for the spin direction of nucleon $1(2)$, the probability to form a spin-singlet state is
\begin{eqnarray}
\left| \left< \chi _{0,0} \mid \Psi \right> \right|^2 = \frac{1}{4}\left[ 1-\cos \theta _1\cos \theta _2-\sin \theta _1\sin \theta _2\cos \left( \phi _1-\phi _2 \right) \right], \notag
\end{eqnarray}
with $\chi _{0,0}$ and $\Psi$ defined in Appendix \ref{app}. The probability to form a single-triplet state is thus $1-\left| \left< \chi _{0,0} \mid \Psi \right> \right|^2$. We have also used the spin- and isospin-dependent Pauli blocking, with the local phase-space cell more specific for nucleons with different spin and isospin states. While isospin has only two states, the expectation direction of the spin can be of any direction ($\phi$, $\theta$). We count the number of nucleons with their spins in each bin ($\Delta\phi$, $\Delta\theta$) in the solid angle, and the occupation probability for a certain phase-space cell with a particular spin state is then evaluated from the projection contribution of all bins.

How the spins of nucleons change after a successful collision is largely unknown. In the present study, we assume that during the collision the colliding nucleons interact with each other through the short-range spin-orbit interaction $v_{LS}=v^r_{so}(r) \vec{L}^{CM}_r \cdot \vec{S}$, where $\vec{S}=\vec{\sigma}_1+\vec{\sigma}_2$ is the total spin of the colliding nucleons, $\vec{L}^{CM}_r$ is the orbital angular momentum of the colliding nucleons with respective to the center of mass (C.M.) of the collision. It is known that the zero-range Skyrme-type spin-orbit interaction is a reduced form of the finite-range spin-orbit interaction~\cite{Bell1956-BELCTN,Vautherin:1971aw}. Here we take $v^r_{so}(r)$ as that fitted by proton-proton scatterings~\cite{Wiringa:1994wb}, i.e.,
\begin{equation}
v^r_{so}\left( r \right) =I T_{\mu}^{2}\left( r \right) +\left[ P+Q(\mu r)+R\left( \mu r \right) ^2 \right] W\left( r \right),
\end{equation}
with
\begin{equation}
T_{\mu}\left( r \right) =\left[ 1+\frac{3}{\mu r}+\frac{3}{\left( \mu r \right) ^2} \right] \frac{e^{-\mu r}}{\mu r}\left( 1-e^{-cr^2} \right) ^2,
\end{equation}
and
\begin{equation}
W\left( r \right) =\left[ 1+e^{\left( r-r_0 \right) /a} \right] ^{-1}.
\end{equation}
In the above, $\mu$ is the pion mass, and $a=0.2$ fm, $r_0=0.5$ fm, and $c=2.1$ fm$^{-2}$ are the shape parameters. For the spin-singlet state ($S=0$) of nucleon pairs, $v^r_{so}(r)$ vanishes. For the spin-triplet state ($S=1$) of nucleon pairs, the values of $I$, $P$, $Q$, and $R$ for different isospin states ($T$) are given in Table~\ref{T1}. While the above short-range spin-orbit interaction is that in vacuum, the corresponding in-medium interaction is expected to have the similar magnitude, by comparing, e.g., the spin-dependent cross sections from nuclear interaction in free space and that in nuclear medium based on the Dirac-Brueckner approach~\cite{PhysRevC.61.014309}.

\begin{table}\small
\caption{Parameters of the finite-range spin-orbit interaction in MeV for different isospin states ($T$) of nucleon pairs.}
\begin{tabular}{|c|c|c|c|c|}
\hline
$T$&$I$&$P$&$Q$&$R$\\ \hline
1&-0.62697&-570.5571&-309.3605&819.1222\\ \hline
0&0.10180&86.0658&46.6655&-356.5175\\ \hline
    \end{tabular}
  \label{T1}
\end{table}

In each time step for the collision process, the short-range spin-orbit interaction may lead to the procession of the nucleon spin according to
\begin{equation}
\frac{d}{dt}\vec{\sigma}_{1,2}=v^r_{so}\left( r \right) \vec{\sigma}_{1,2} \times \vec{L}^{CM}_r.
\end{equation}
In the present study with spin degree of freedom, $\vec{L}^{CM}_r$ is not a conserved quantity, while the total angular momentum $\vec{L}^{CM}=\vec{L}^{CM}_r+\vec{S}$ is conserved. Defining $\Delta \vec{\sigma}=\vec{\sigma}_1-\vec{\sigma}_2$, the above equations can be rewritten as
\begin{eqnarray}\label{procc}
\frac{d}{dt}\vec{S}&=&v^r_{so}\left( r \right)\vec{S} \times \vec{L}^{CM},\\
\frac{d}{dt}\Delta\vec{\sigma}&=&v^r_{so}\left( r \right)\Delta\vec{\sigma} \times (\vec{L}^{CM}-\vec{S}).
\end{eqnarray}
The above coupled equations can be solved numerically. Actually, it turns out that the spin procession angle from the above equations is rather small, which means that the spin direction is largely unchanged after a successful nucleon-nucleon collision.

\subsection{Angular momentum conservation for each collision}

The angular momentum conservation for nucleon-nucleon collisions in transport simulation was first investigated in Ref.~\cite{Gale:1990zz} and recently revisited in Ref.~\cite{Liu:2023pgc}. Without considering nucleon spin, the conservation of the orbital angular momentum requires in-plane nucleon-nucleon collisions. While there are different prescriptions for such in-plane collisions, considerable effects on the overall dynamics due to the constraint of angular momentum conservation have been observed~\cite{Liu:2023pgc}. In the following, we present an improved treatment in the presence of nucleon spins compared to that discussed in Ref.~\cite{Liu:2023pgc}.

The total angular momentum conservation for a collision between nucleon 1 and nucleon 2 with coordinates $\vec{r}_{1(2)}$, momenta $\vec{p}_{1(2)}$, and spins $\vec{\sigma}_{1(2)}$ requires
\begin{equation}\label{amc}
\vec{r}_1\times \vec{p}_1+\vec{r}_2\times \vec{p}_2+\vec{\sigma}_1+\vec{\sigma}_2=\vec{r}_{1}^{'}\times \vec{p}_{1}^{'}+\vec{r}_{2}^{'}\times \vec{p}_{2}^{'}+\vec{\sigma}_{1}^{'}+\vec{\sigma}_{2}^{'}.
\end{equation}
From now on, we take the convention that quantities with (without) ``$\prime$'' represent those after (before) the nucleon-nucleon collision, and quantities with (without) an asterisk represent those in the C.M. (lab) frame of the collision. Defining the central and relative coordinates and momenta as
\begin{eqnarray}
&&\vec{R}=(\vec{r}_{1}+\vec{r}_{2})/2, \vec{r}=(\vec{r}_{1}-\vec{r}_{2})/2, \notag\\
&&\vec{P}=\vec{p}_1+\vec{p}_2, \vec{p}=\vec{p}_{1}-\vec{p}_{2}, \notag
\end{eqnarray}
Eq.~(\ref{amc}) is identical to
\begin{align*}
\vec{R}\times \vec{P}+\vec{r}\times \vec{p}+\vec{\sigma}_1+\vec{\sigma}_2=\vec{R}^{'}\times \vec{P}^{'}+\vec{r}^{'}\times \vec{p}^{'}+\vec{\sigma}_{1}^{'}+\vec{\sigma}_{2}^{'}.
\end{align*}

Since $\vec{R}=\vec{R}^{'}$ and $\vec{P}=\vec{P}^{'}$ are satisfied, the task is to find proper $\vec{r}^{'}$ and $\vec{p}^{'}$ which lead to a given orbital angular momentum $\vec{L}_{r}^{'} = \vec{r}^{'} \times \vec{p}^{'}$ once $\vec{\sigma}_{1}^{'}+\vec{\sigma}_{2}^{'}$ is determined. To do that, first we need to select properly the direction of the C.M. momentum $\vec{p}_1^{*'}$ in such a way that the relative momentum $\vec{p}^{'}$ is perpendicular to the angular momentum $\vec{L}_{r}^{'}$, i.e., $\vec{p}^{'} \cdot \vec{L}_{r}^{'} = 0$. Through Lorentz transformation, $\vec{p}^{'} \cdot \vec{L}_{r}^{'}$ can be further expressed as
\begin{align*}
\vec{p}^{'} \cdot \vec{L}_{r}^{'} ={\vec{p}_1^{*'}}\cdot \left( 2\vec{L}_{r}^{'}+\frac{2\gamma ^2}{\gamma +1}\vec{\beta }\cdot \vec{L}_{r}^{'}\vec{\beta} \right) +\gamma\Delta e^{*}\vec{\beta }\cdot \vec{L}_{r}^{'}.
\end{align*}
In the above, $\vec{\beta}$ is the velocity of the C.M. frame of the nucleon-nucleon collision with respect to the lab frame, $\gamma = 1/\sqrt{1-\beta^2}$ is the Lorentz factor, and $\Delta e^*=\sqrt{{{p}_{1}^{*'}}^2+{m}_{1}^{2}}-\sqrt{{{p}_{2}^{*'}}^2+{m}_{2}^{2}}$ is the energy difference between two particles after collisions in the C.M. frame, which vanishes for elastic collisions with $m_1=m_2=m$. The above equation shows that the direction of $\vec{p}_1^{*'}$ should be perpendicular to
\begin{equation}
\vec{q} = 2\vec{L}_{r}^{'}+\frac{2\gamma ^2}{\gamma +1}\vec{\beta }\cdot \vec{L}_{r}^{'}\vec{\beta}.
\end{equation}
In the collision treatment, we first determine the polar angle $\theta$ according to the differential cross section. Afterwards, we select the direction of $\vec{p}_1^{*'}$  perpendicular to $\vec{q}$ by setting
\begin{equation}
{\vec{p}_1^{*'}}={p_1^{*'}}\left( \cos \varphi \vec{\hat{e}}_1+\sin \varphi \vec{\hat{e}}_2 \right), \label{p1s}
\end{equation}
with the unit vectors $\vec{\hat{e}}_1$ and $\vec{\hat{e}}_2$ written as
\begin{eqnarray}
\vec{\hat{e}}_1 = \frac{\vec{q}\times \vec{p}_1^{*}}{\left| \vec{q}\times \vec{p}_1^{*} \right|},~~
\vec{\hat{e}}_2=\frac{\vec{q}\times \vec{p}_1^{*}\times \vec{q}}{\left| \vec{q}\times \vec{p}_1^{*}\times \vec{q} \right|}.
\end{eqnarray}
It is easy to see that $\vec{\hat{e}}_1$ and $\vec{\hat{e}}_2$ form a plane perpendicular to $\vec{q}$, and the angle $\varphi$ characterizes the direction of $\vec{p}_1^{*'}$ in that plane. By substituting Eq.~(\ref{p1s}) into ${\vec{p}_1^{*'}}\cdot \vec{p}_1^{*}={p_1^{*'}} p_1^{*}\cos \theta$, we get
\begin{equation}
\sin \varphi =\frac{\cos \theta}{\vec{\hat{e}}_2\cdot \vec{p}_1^{*}}.
\end{equation}
Since we have the constraint $|\sin \varphi| \le 1$, the range of $\cos\theta$ is also limited to $\left| \vec{\hat{e}}_2\cdot \vec{p}_1^{*} \right|\ge \left| \cos \theta \right|$, and this is equivalent to
\begin{equation}
-\frac{\left| \vec{p}_1^{*}\times \vec{q} \right| }{q p_1^{*}}\le \cos \theta \le \frac{\left| \vec{p}_1^{*}\times \vec{q} \right| }{q p_1^{*}}.
\end{equation}

Now the direction of $\vec{r}^{'}\times \vec{p}^{'}$ is the same as $\vec{L}_{r}^{'}$. To ensure that they have the same magnitude, the relative coordinates $\vec{r}^{'}$ of the colliding nucleons must satisfy
\begin{equation}
\vec{r}^{'}=  A \vec{\hat{e}}_{\vec{p}^{'}}+\frac{L_{r}^{'}}{p^{'}}\vec{\hat{e}}_{\vec{p}^{'}\times \vec{L}_{r}^{'}} ,
\end{equation}
where $\vec{\hat{e}}_{\vec{p}^{'}}$ and $\vec{\hat{e}}_{\vec{p}^{'}\times \vec{L}_{r}^{'}}$ are the unit vector in the direction of $\vec{p}^{'}$ and $\vec{p}^{'}\times \vec{L}_{r}^{'}$, respectively. $A$ can be any real number, and its value is determined by minimizing $|\vec{r}^{'}-\vec{r}|$, which leads to~\cite{Gale:1990zz}
\begin{equation}
\vec{r}^{'}=\left( \vec{r}\cdot \vec{\hat{e}}_{\vec{p}^{'}} \right) \vec{\hat{e}}_{\vec{p}^{'}}+\frac{L_{r}^{'}}{p^{'}}\vec{\hat{e}}_{\vec{p}^{'}\times \vec{L}_{r}^{'}}.
\end{equation}
It can be verified that the total angular momentum is now conserved with a minimum $|\vec{r}^{'}-\vec{r}|$.

\subsection{Angular momentum conservation for mean-field evolution}

Besides nucleon-nucleon collisions, the mean-field evolution may also violate the conservation of the total angular momentum, which includes the contribution from the orbital angular momentum and the nucleon spin. To correct for that, we use the following method in the similar spirit as in Ref.~\cite{Papa:2005sp}. At each time step, we correct the momentum of each particle with a small amount $\Delta \vec{p}_i$ to ensure the conservation of the total angular momentum, with its each component expressed as
\begin{equation}
\Delta J_c=\sum_i\varepsilon_{abc}r_{i,a}\Delta p_{i,b},
\end{equation}
where $\varepsilon_{abc}$ is the Levi-Civita symbol, $a$, $b$, and $c$ represent symbols of the Cartesian coordinate, and double symbols imply summation. To make the global correction small enough, we define a function
\begin{equation}
\mathcal{L}=\sum_i\Delta p_{i,a}\Delta p_{i,a}+\lambda_c\left( \sum_i \varepsilon _{abc}r_{i,a}\Delta p_{i,b}-\Delta J_c \right),
\end{equation}
and the values of $\Delta p_{i,a}$ and $\lambda_c$ can be obtained by taking the partial derivatives of this function
$$
\left\{ \begin{array}{l}\frac{\partial \mathcal{L}}{\partial \Delta p_{i,a}}=0\\ \frac{\partial \mathcal{L}}{\partial \lambda _c}=0\\ \end{array} \right. \Rightarrow \left\{ \begin{array}{l}\Delta p_{i,a}=\frac{1}{2}\varepsilon _{abc}r_{i,b}\lambda _{c}\\ \lambda _{c} \sum_i r_{i,a}r_{i,a}-\lambda_{a}\sum_i r_{i,a}r_{i,c}+2\Delta J_{c}=0\\ \end{array} \right..
$$
In this way, the correction of the momentum $\Delta p_{i,a}$ can be obtained once we solve a ternary system of equations about $\lambda_c$, which can help us to conserve the total angular momentum in the mean-field evolution. The resulting $\Delta p_{i}$ is about a few keV/c for each nucleon. Since there are many test particles used in the study, we note that the violations of the total energy and the total momentum are of the order $0.0001\%$, smaller than the numerical fluctuation.

\subsection{Spin-dependent coalescence model}
\label{scluster}

We use a spin-dependent coalescence model to study the production of light clusters with different spins, with minor improvement compared to that in Ref.~\cite{Xia:2014rua}. The yields of deuterons ($d$) as well as tritons ($t$) and $^3$He from free nucleons are expressed by the following Wigner function forms~\cite{Chen:2003ava,Sun:2017ooe}
\begin{eqnarray}
f_d &=& 8g_d\exp \left( -\frac{\rho ^2}{\sigma _{d}^{2}}-p_{\rho}^{2}\sigma _{d}^{2} \right), \\
f_{t/^3He} &=& 8^2g_{t/^3He}\exp \left( -\frac{\rho ^2+\lambda ^2}{\sigma _{t/^3He}^{2}}-\left( p_{\rho}^{2}+p_{\lambda}^{2} \right) \sigma _{t/^3He}^{2} \right).
\end{eqnarray}
In the above,
\begin{eqnarray}
&\vec{\rho} =\frac{1}{\sqrt{2}}(\vec{r}_{1}-\vec{r}_{2}),
\vec{p}_{\rho} =\frac{1}{\sqrt{2}} (\vec{p}_{1}-\vec{p}_{2}), \notag\\
&\vec{\lambda} =\frac{1}{\sqrt{6}}\left(\vec{r}_{1}+\vec{r}_{2}-2\vec{r}_{3}\right),
\vec{p}_{\lambda} =\frac{1}{\sqrt{6}} (\vec{p}_{1}+\vec{p}_{2}-2\vec{p}_{3}) \notag
\end{eqnarray}
are relative coordinates and momenta, and $\sigma _{d}=2.26$ fm, $\sigma _{t}=1.59$ fm, and $\sigma _{^3He}=1.76$ fm are the width parameters for the Wigner functions that reproduce the root-mean-square radii of corresponding light clusters~\cite{Ropke:2008qk}. In the spin-averaged case, the statistical factors have the values $g_d=3/4$ and $g_{t/^3He}=1/4$. With the information of nucleon spin explicitly available, the statistic factor $g_d$ for deuterons with a neutron and a proton forming a spin-triplet state can be calculated from
\begin{eqnarray}
g_d &=& \left| \left< \chi _{1,1} \mid \Psi \right> \right|^2 + \left| \left< \chi _{1,0} \mid \Psi \right> \right|^2 +\left| \left< \chi _{1,-1} \mid \Psi \right> \right|^2 \notag\\
&=& 1-\left| \left< \chi _{0,0} \mid \Psi \right> \right|^2 \notag\\
&=& \frac{1}{4}\left[ 3+\cos \theta _1\cos \theta _2+\sin \theta _1\sin \theta _2\cos \left( \phi _1-\phi _2 \right) \right], \notag\\
\end{eqnarray}
where $\theta_{1(2)}$ and $\phi_{1(2)}$ are the azimuthal angle and polar angle for the spin direction of nucleon 1(2), and the definitions of $\chi _{1,1}$, $\chi _{1,0}$, $\chi _{1,-1}$, $\chi _{0,0}$, and $\Psi$ can be found in Appendix \ref{app}.  The $\rho_{0,0}$ term of the spin arrangement matrix characterizing the spin alignment can be written as
\begin{eqnarray}
\rho _{0,0}&=&\frac{\left| \left< \chi _{1,0} \mid \Psi \right> \right|^2}{1-\left| \left< \chi _{0,0} \mid \Psi \right> \right|^2} \notag\\
&=&\frac{1-\cos \theta _1\cos \theta _2+\sin \theta _1\sin \theta _2\cos \left( \phi _1-\phi _2 \right)}{3+\cos \theta _1\cos \theta _2+\sin \theta _1\sin \theta _2\cos \left( \phi _1-\phi _2 \right)},
\end{eqnarray}
where the $\sin \theta _1\sin \theta _2\cos \left( \phi _1-\phi _2 \right)$ term is a small correction term compared to that in Ref.~\cite{Liang:2004xn}. The spin state of tritons ($^3$He) is determined by the additional neutron (proton), and the rest neutron-proton pair form a spin-singlet state, so the statistic factor can be calculated from
\begin{eqnarray}
g_{t/^3He}&=&\left| \left< \chi _{0,0}\mid \Psi \right> \right|^2 \notag\\
&=&\frac{1}{4}\left[ 1-\cos \theta _1\cos \theta _2-\sin \theta _1\sin \theta _2\cos \left( \phi _1-\phi _2 \right) \right]. \notag\\
\end{eqnarray}

\section{Results and discussions}
\label{results}

We explore the spin dynamics in intermediate-energy heavy-ion collisions based on the improved SIBUU transport model, and compare extensively the spin-dependent observables with and without the constraint of rigorous angular momentum conservation (AMC). These spin-dependent observables include the global and local spin polarizations as well as transverse and elliptic flows for nucleons and light clusters of different spin states. Compared with traditional IBUU calculations, we will also illustrate the effect of the spin-dependent potential on the spin-averaged transverse flow. According to our previous studies, the spin polarization as well as the spin splitting of the collective flow are largest at the beam energy of around 100 AMeV~\cite{Xia:2014qva,Xia:2019whr}. To have an appreciable angular momentum and bulk medium to illustrate the effect of the spin-orbit potential, the following discussions will be mainly focused on midcentral Au+Au collisions at the beam energy of 100 AMeV.

\subsection{Spin dynamics and polarization}

\begin{figure*}[ht]
\includegraphics[width=0.5\linewidth]{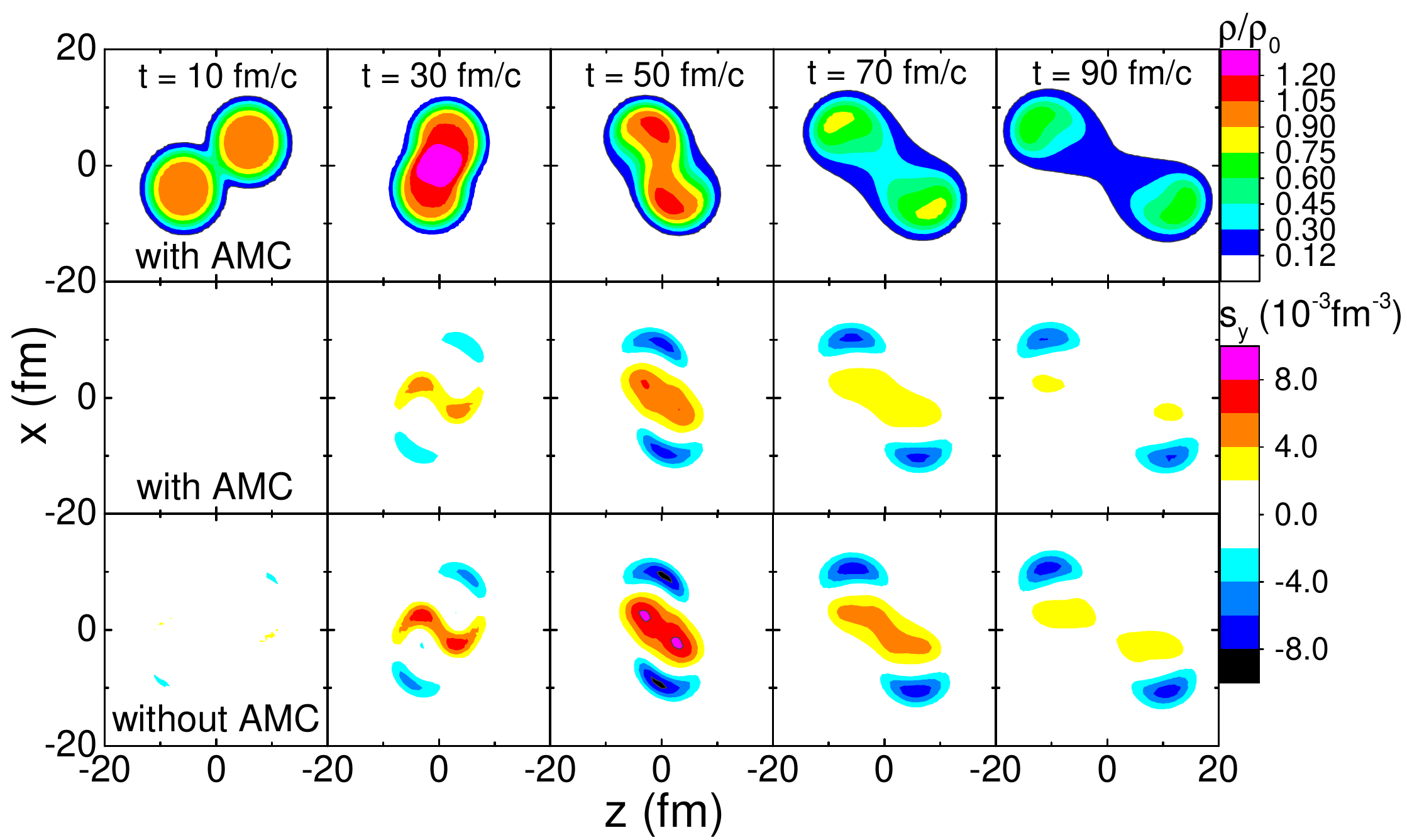}
\includegraphics[width=0.35\linewidth]{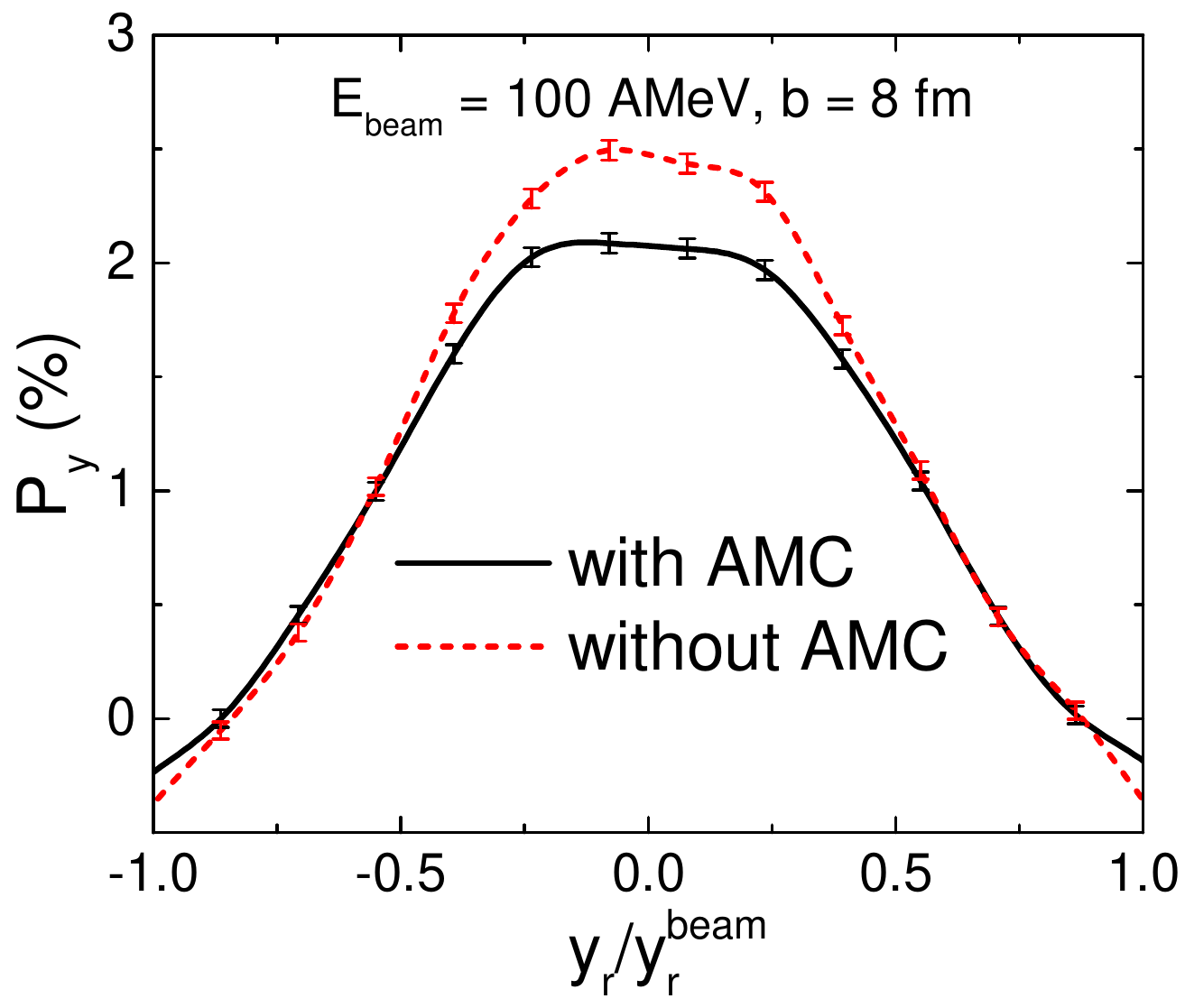}
\caption{\label{den_xoz} Left: Contours of the reduced density $\rho/\rho_0$ (first row), and the $y$ component of the spin density with (second row) and without (third row) angular momentum conservation in the x-o-z plane, in Au+Au collisions at the beam energy of 100 AMeV and impact parameter $\text{b}=8$ fm. Right: Spin polarization of free nucleons in $y$ direction as a function of reduced rapidity $y_r/y_r^{beam}$ in the same reaction of the left window.}
\end{figure*}

\begin{figure*}[ht]
\includegraphics[width=0.5\linewidth]{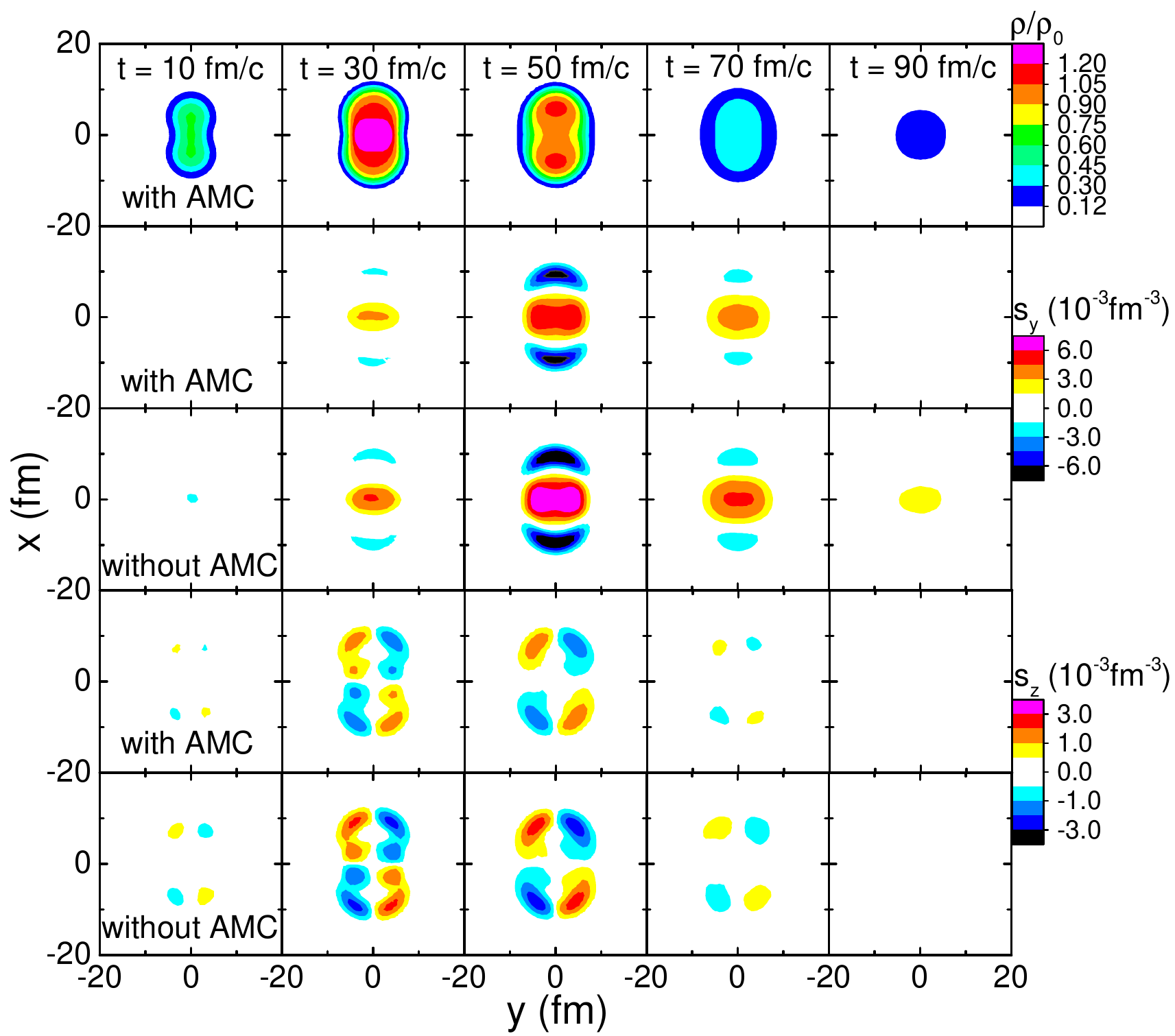}
\includegraphics[width=0.32\linewidth]{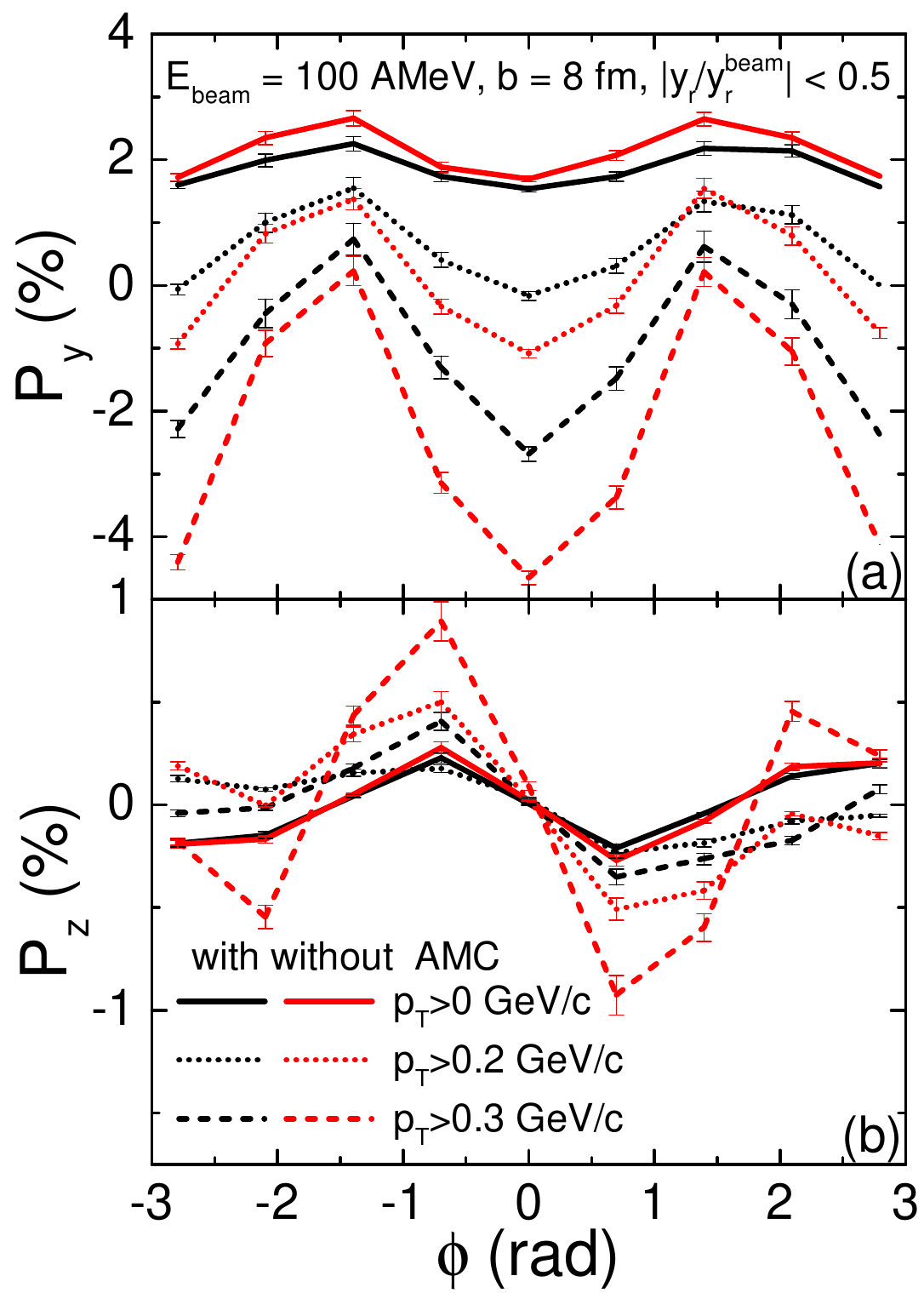}
\caption{\label{den_xoy} Left: Contours of the reduced density $\rho/\rho_0$ (first row), the $y$ component of the spin density with (second row) and without (third row) angular momentum conservation, and the $z$ component of the spin density with (fourth row) and without (fifth row) angular momentum conservation in the x-o-y plane, in Au+Au collisions at the beam energy of 100 AMeV and impact parameter $\text{b}=8$ fm. Right: Spin polarization of mid-rapidity free nucleons at different transverse momenta $p_T$ in $y$ (a) and $z$ (b) directions as a function of azimuthal angle $\phi$ in the same reaction of the left window.}
\end{figure*}

\begin{figure}[ht]
\includegraphics[width=0.8\linewidth]{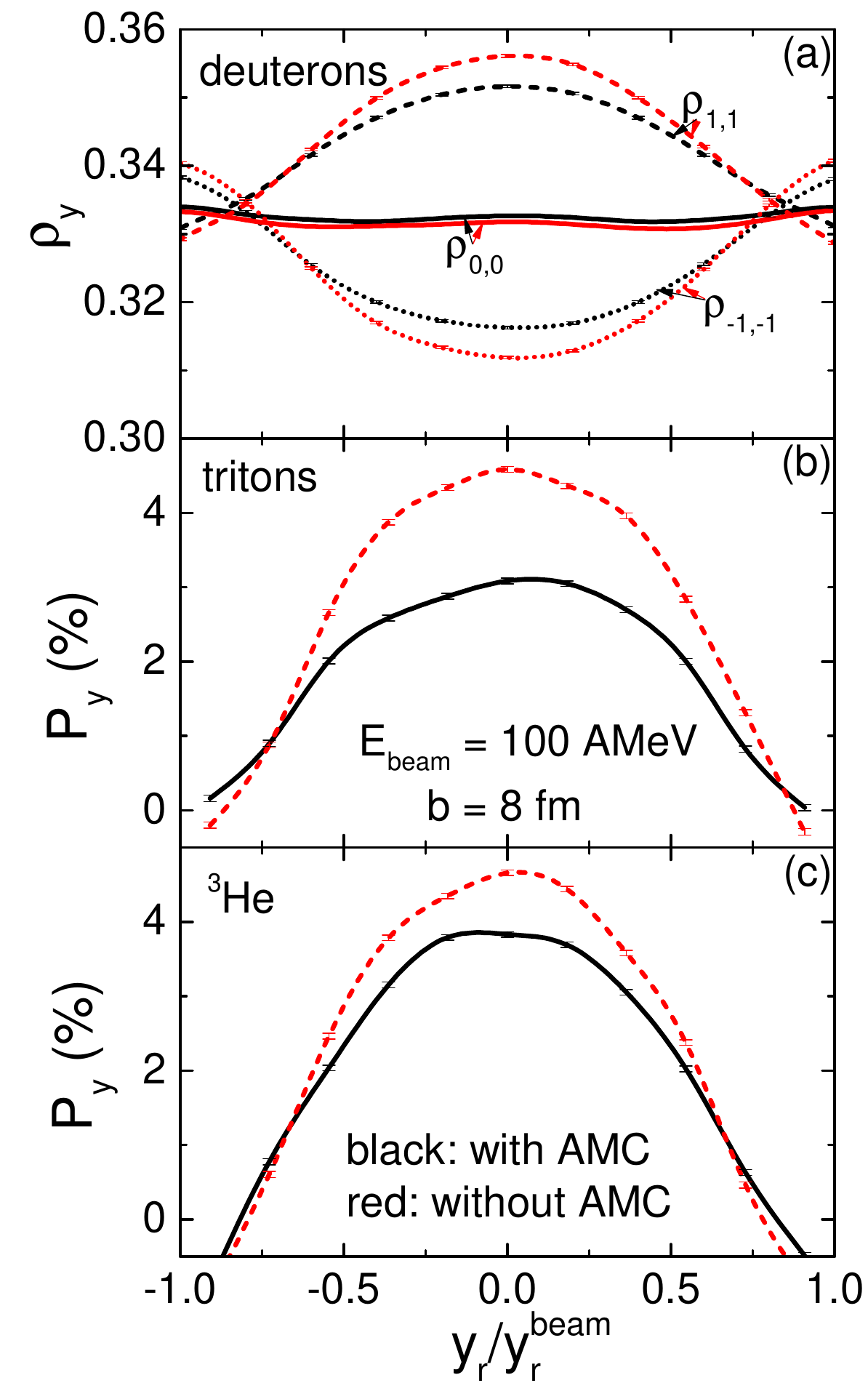}
\caption{\label{Polcl} Triangular components of the spin arrangement matrix for deuterons (a) as well as spin polarization of tritons (b) and $^3$He (c) in $y$ direction as a function of reduced rapidity $y_r/y_r^{beam}$ in Au+Au collisions at the beam energy of 100 AMeV and impact parameter $\text{b}=8$ fm with and without angular momentum conservation.}
\end{figure}

\begin{figure}[ht]
\includegraphics[width=0.8\linewidth]{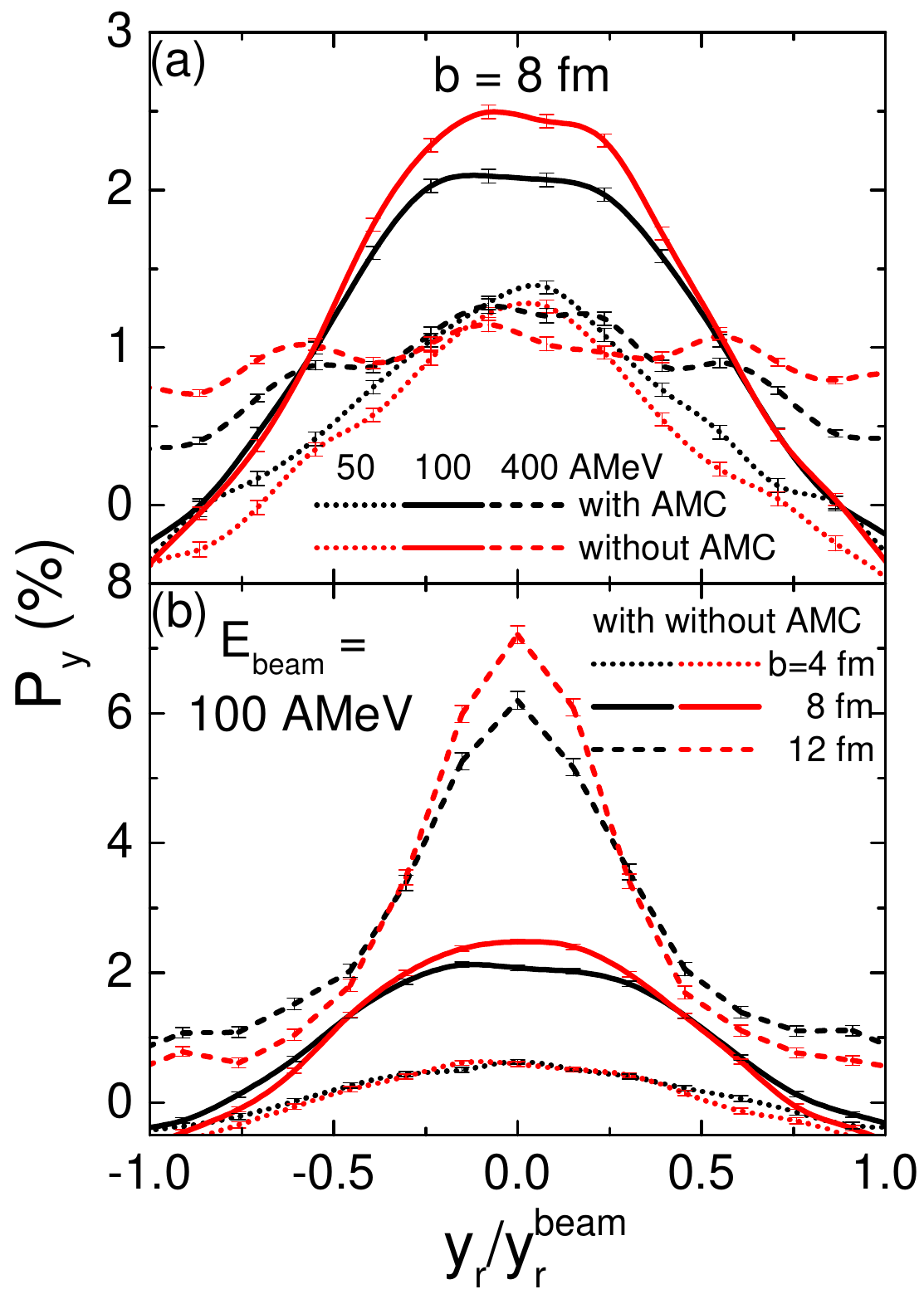}
\caption{\label{Polnu} Comparison of the spin polarization of free nucleons in $y$ direction as a function of reduced rapidity $y_r/y_r^{beam}$ in Au+Au collisions at different beam energies but a fixed impact parameter $\text{b}=8$ fm (a) and at a fixed beam energy of 100 AMeV and different impact parameters $\text{b}=4$, 8, and 12 fm (b) with and without angular momentum conservation.}
\end{figure}

We begin with showing contours of nucleon number density and spin density in the reaction plane (x-o-z plane) in the left window of Fig.~\ref{den_xoz}, for the typical midcentral Au+Au collisions at $E_{lab}=100$ AMeV. Here we only show the density evolution with AMC, which leads to a slightly lower density compared to the calculation without AMC (see Fig.~5 of Ref.~\cite{Liu:2023pgc}). Generally, the global spin polarization, shown by the $y$ component of the spin density $s_y$, is generated by the dominating contribution of the time-odd term $(\nabla \times \vec{j})_y$ over that of the time-even term $(\nabla \rho \times \vec{p})_y$, as demonstrated in Ref.~\cite{Xia:2019whr}. The lower density with AMC thus generally leads to a weaker $s_y$ in both participant and spectator regions. The right window of Fig.~\ref{den_xoz} displays the rapidity distribution of the global spin polarization of free nucleons, with the latter identified by their low densities $\rho<\rho_0/8$ in SIBUU simulations. The global spin polarization in $y$ direction perpendicular to the reaction plane is defined as
\begin{equation}\label{py}
P_y = \frac{N_{\mathcal{s}_y=+\frac{1}{2}}-N_{\mathcal{s}_y=-\frac{1}{2}}}{N_{\mathcal{s}_y=+\frac{1}{2}}+N_{\mathcal{s}_y=-\frac{1}{2}}},
\end{equation}
where $N_{\mathcal{s}_y=\pm\frac{1}{2}}$ is the number of free nucleons at spin state $\mathcal{s}_y = \pm\frac{1}{2}$. Such spin polarization was measured for $\Lambda$ hyperons in relativistic heavy-ion collisions~\cite{STAR:2017ckg}, and here we are discussing the spin polarization for free nucleons in intermediate-energy heavy-ion collisions. The global spin polarization is generally stronger at midrapidities, where it is dominated by the spin density $s_y$ of the participant, and weaker at large rapidities, where it is affected by the spin density $s_y$ of the spectators. It is seen that SIBUU with AMC predicts a slightly weaker global spin polarization compared to that without AMC. This is consistent with the weaker $s_y$ with the constraint of AMC shown in the left window of Fig.~\ref{den_xoz}.

The density contours in the transverse plane (x-o-y plane) may provide us with additional information of the spin dynamics in a different point of view, and they are displayed in the left window of Fig.~\ref{den_xoy}. While the general features of the evolution of nucleon number density as well as the $y$ and $z$ components of the spin densities $s_y$ and $s_z$ respectively are similar to those shown in Ref.~\cite{Xia:2019whr}, weaker $s_y$ and $s_z$ are observed with AMC compared to that without AMC in both participant and spectator regions. This is again due to the slightly lower density reached with AMC. We also display the azimuthal angular distribution of the spin polarization for free nucleons in $y$ and $z$ directions in the right window of Fig.~\ref{den_xoy}, with the azimuthal angle calculated from $\phi = \text{atan2}(p_y,p_x)$. Here $\phi=0$ and $\pm \pi$ correspond to nucleons that emit in $\pm x$ directions, and $\phi=\pm \frac{\pi}{2}$ corresponds to nucleons that emit in $\pm y$ directions. It is seen from panel (a) that the $P_y$ of total free nucleons, which include those from both participant and spectator regions, has a flat $\phi$ distribution. The $\phi$ dependence becomes stronger for nucleons with larger transverse momenta $p_T$, since they emit at early stages and their $P_y$ reflect the different $s_y$ in the participant and spectator regions. It is seen that the $P_y$ are affected differently by the constraint of AMC for low- and high-$p_T$ nucleons. The spin polarization $P_z$ in $z$ direction, which is defined similarly as Eq.~(\ref{py}), is called the local spin polarization, and for $\Lambda$ hyperons its azimuthal angular dependence measured by STAR Collaboration~\cite{PhysRevLett.123.132301} has attracted considerable attentions~\cite{Becattini:2021suc,Fu:2021pok,PhysRevLett.120.012302,PhysRevLett.125.062301}. Without the constraint of AMC, the $\phi$ distribution of $P_z$ is consistent with $s_z$ shown in the left window, and it has an opposite sign compared to that for $\Lambda$ hyperons at relativistic energies~\cite{PhysRevLett.123.132301}. With the constraint of AMC, while the $\phi$ dependence is qualitatively similar, the $P_z$ has a smaller magnitude.

The spin polarization of nucleons may affect the production of nuclear clusters in different spin states, and for light clusters this can be studied through a spin-dependent coalescence method as discussed in Sec.~\ref{scluster}. In Fig.~\ref{Polcl} (a), we display the rapidity dependence of the triangular components of the spin arrangement matrix for deuterons, and here $\rho_{1,1}$, $\rho_{0,0}$, and $\rho_{-1,-1}$ correspond to the relative fractions of deuterons in spin states $\mathcal{S}_y=+1$, 0, and $-1$, respectively. It is seen that $\rho_{1,1}$ and $\rho_{-1,-1}$ have oppositive rapidity dependence, while $\rho_{0,0}$ has almost no rapidity dependence and is slightly smaller than $1/3$. The deviation of $\rho_{0,0}$ from $1/3$, which is called the spin alignment, has attracted considerable interest~\cite{Sheng:2019kmk,Sheng:2022wsy}. Referencing the situation here, in relativistic heavy-ion collisions one expects that $\rho_{1,1}$ and $\rho_{-1,-1}$ of vector mesons could be more sensitive to the spin polarization of the quark-gluon plasma, while $\rho_{0,0}$ is more easily measurable~\cite{STAR:2022fan,Liang:2004xn}. For tritons, the neutron-proton pair form a spin-singlet state, so the triton spin is determined by the residue neutron. In Fig.~\ref{Polcl} (b) it is seen that the spin polarization of tritons is similar to that of free nucleons. The result is similar for $^3$He as shown in Fig.~\ref{Polcl} (c), except that the $P_y$ distribution is a little narrower and the $P_y$ with AMC is larger compared to that of tritons, since now the residue proton dominates the spin of $^3$He. In Fig.~\ref{Polcl} one observes a weaker global spin polarization for light clusters with the constraint of AMC, consistent with the weaker global spin polarization of nucleons shown above. For the local spin polarization, it is rather weak for light clusters.

We devote Fig.~\ref{Polnu} to the global spin polarizations for free nucleons at different beam energies and impact parameters. As already observed in Ref.~\cite{Xia:2019whr}, $P_y$ for midrapidity nucleons is larger at about 100 AMeV but smaller at lower or higher energies, and this is due to a weaker vorticity characterized by the $\nabla \times \vec{j}$ term at lower energies, and due to more free nucleons at higher energies. The $P_y$ is also seen to be stronger in more peripheral collisions, since the spin-orbit potential depends on the density gradient and is stronger near the surface of the nuclear matter. While a weaker $s_y$ with AMC compared to that without AMC is observed in each scenario, the resulting $P_y$ of free nucleons may depend on various competition effects. Generally, one sees that the constraint of AMC has some influence on the spin polarizations, while the resulting $P_y$ are still appreciable at midrapidities in most cases. At larger rapidities and especially at smaller beam energies $P_y$ becomes negative, since it is dominated by the spectator matter with a negative $s_y$ (see the left window of Fig.~\ref{den_xoz}). Compared to the results without AMC, the weaker spin-orbit potential with AMC leads to a less negative $s_y$ of the spectator matter, and thus a less negative $P_y$ at larger rapidities especially at smaller beam energies.

\subsection{Spin-dependent flows}

\begin{figure}[ht]
\includegraphics[width=0.8\linewidth]{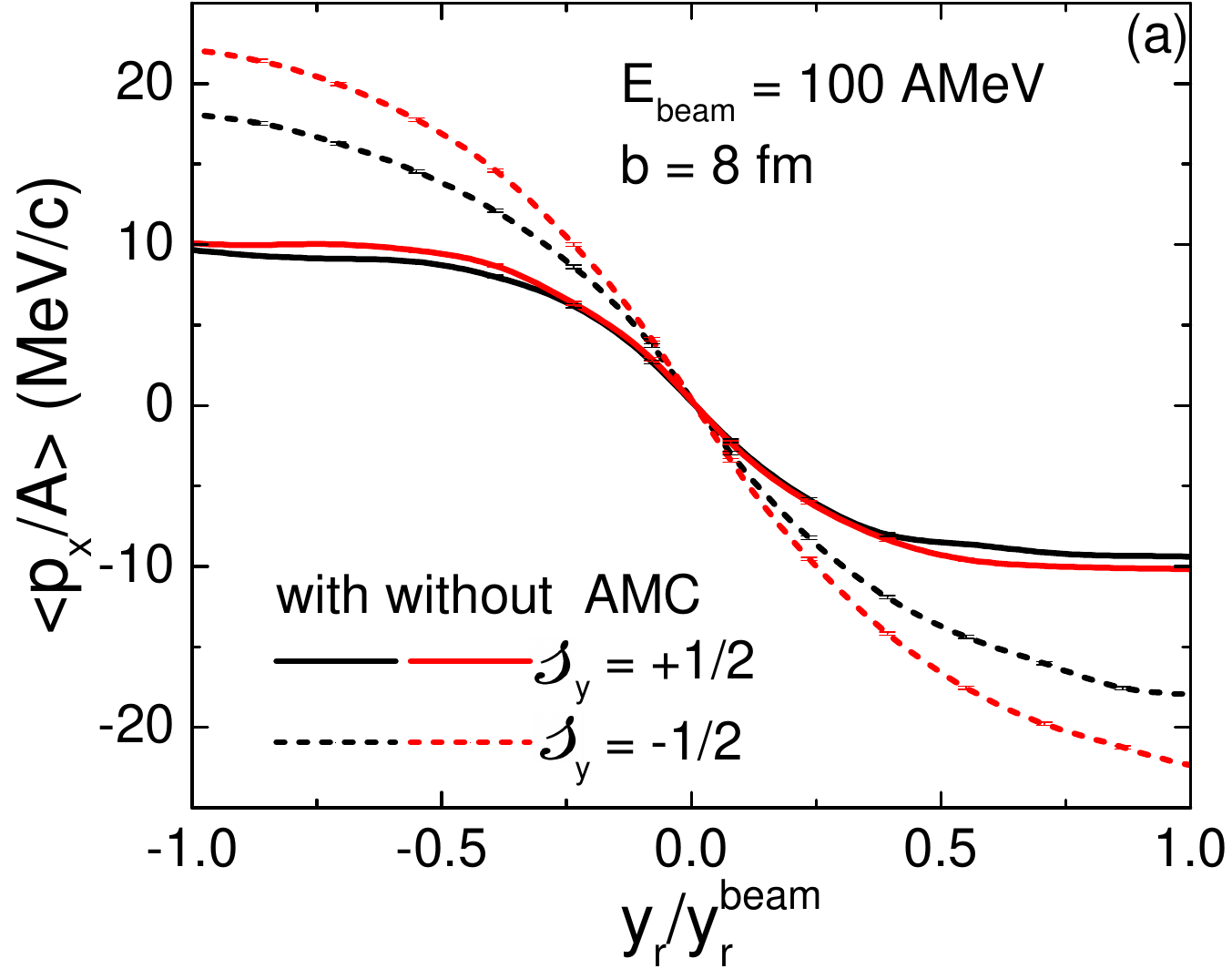}\\
\includegraphics[width=0.8\linewidth]{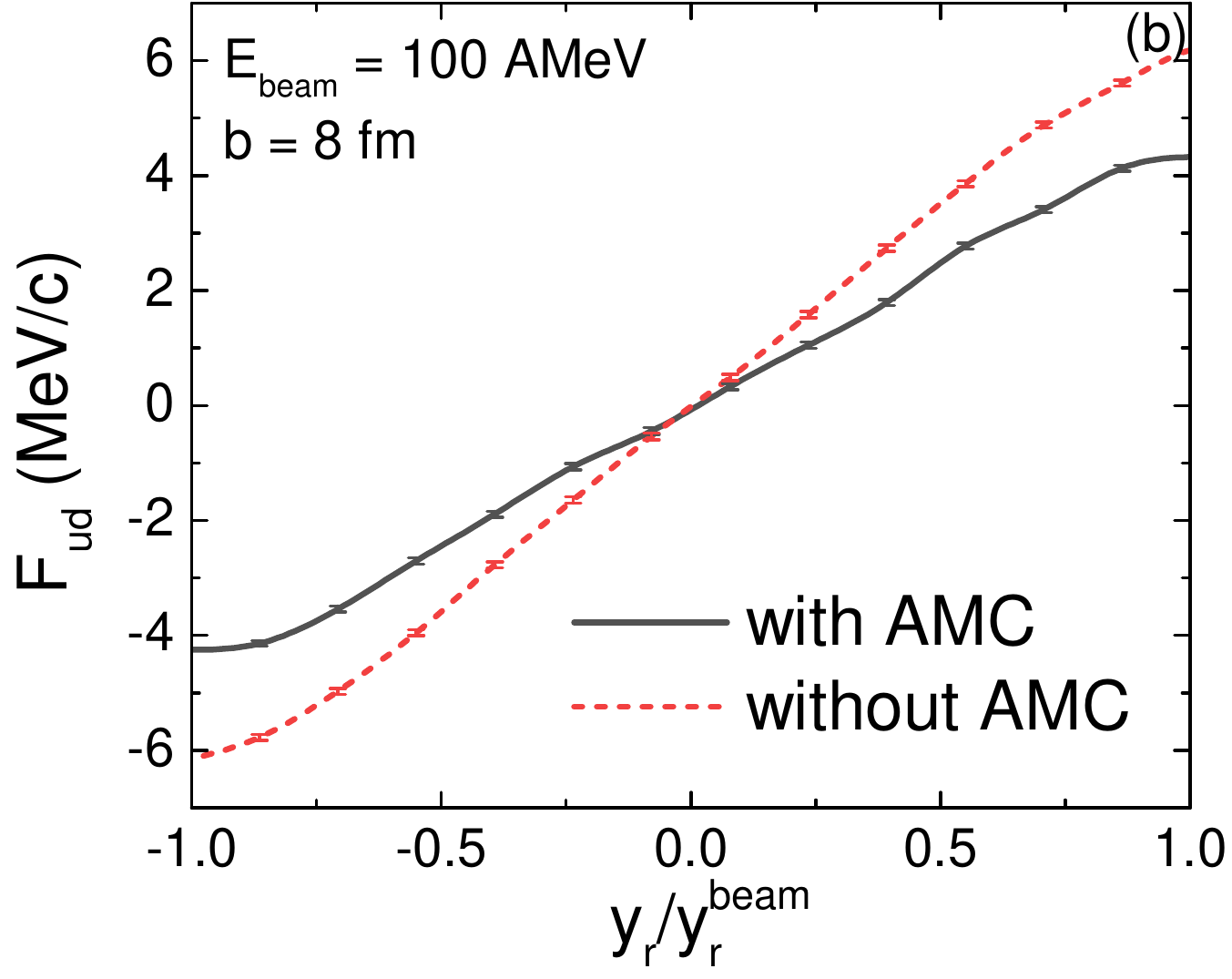}\\
\includegraphics[width=0.75\linewidth]{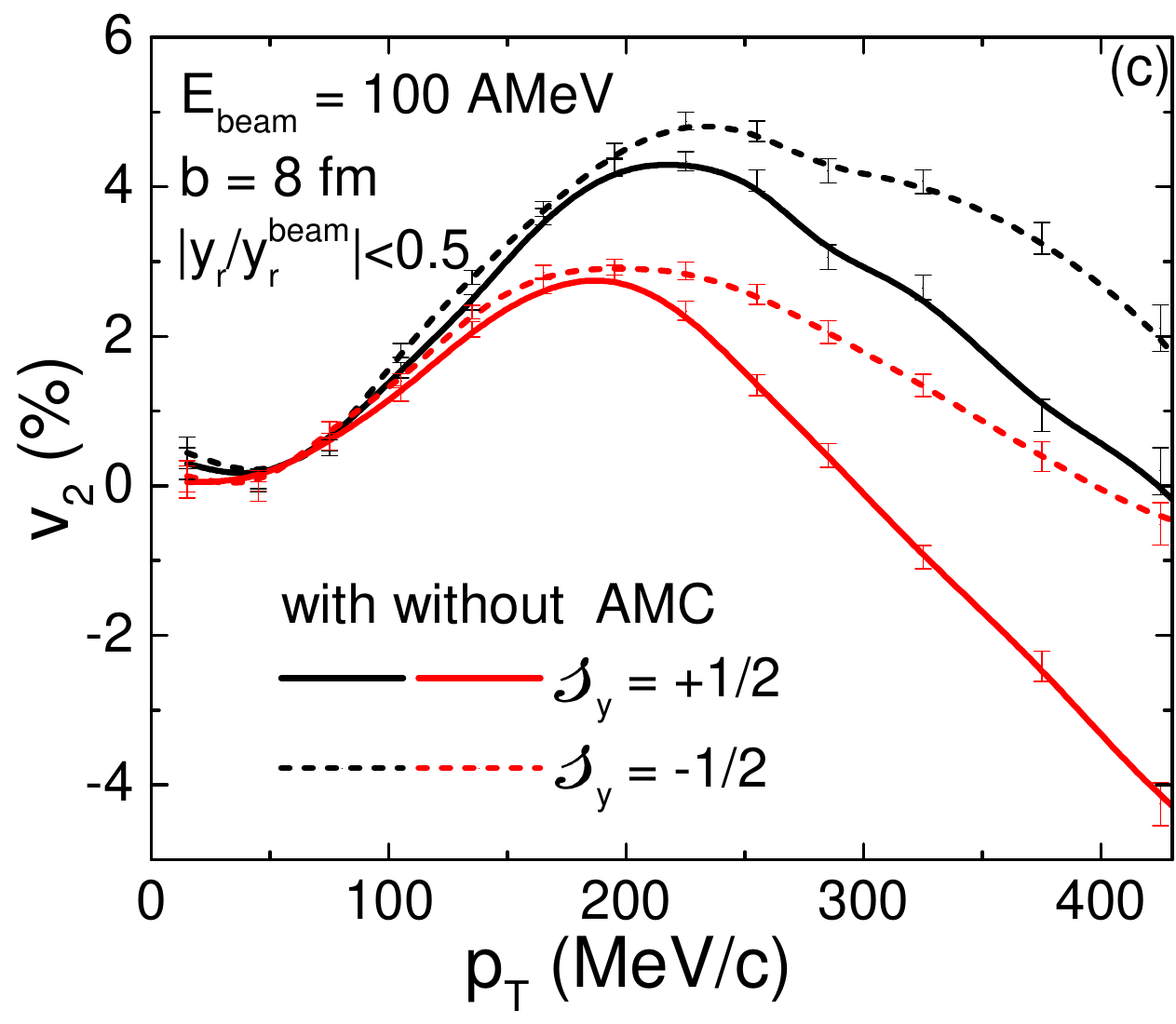}
\caption{\label{px} Transverse flows of spin-up ($\mathcal{s}_y=+\frac{1}{2}$) and spin-down ($\mathcal{s}_y=-\frac{1}{2}$) free nucleons (a) and spin up-down differential transverse flows (b) as a function of reduced rapidity $y_r/y_r^{beam}$, and elliptic flow of spin-up and spin-down free nucleons at midrapidities (c) as a function of transverse momentum $p_T$, in Au+Au collisions at the beam energy of 100 AMeV and impact parameter $\text{b}=8$ fm with and without angular momentum conservation.}
\end{figure}

\begin{figure*}[ht]
\includegraphics[width=0.4\linewidth]{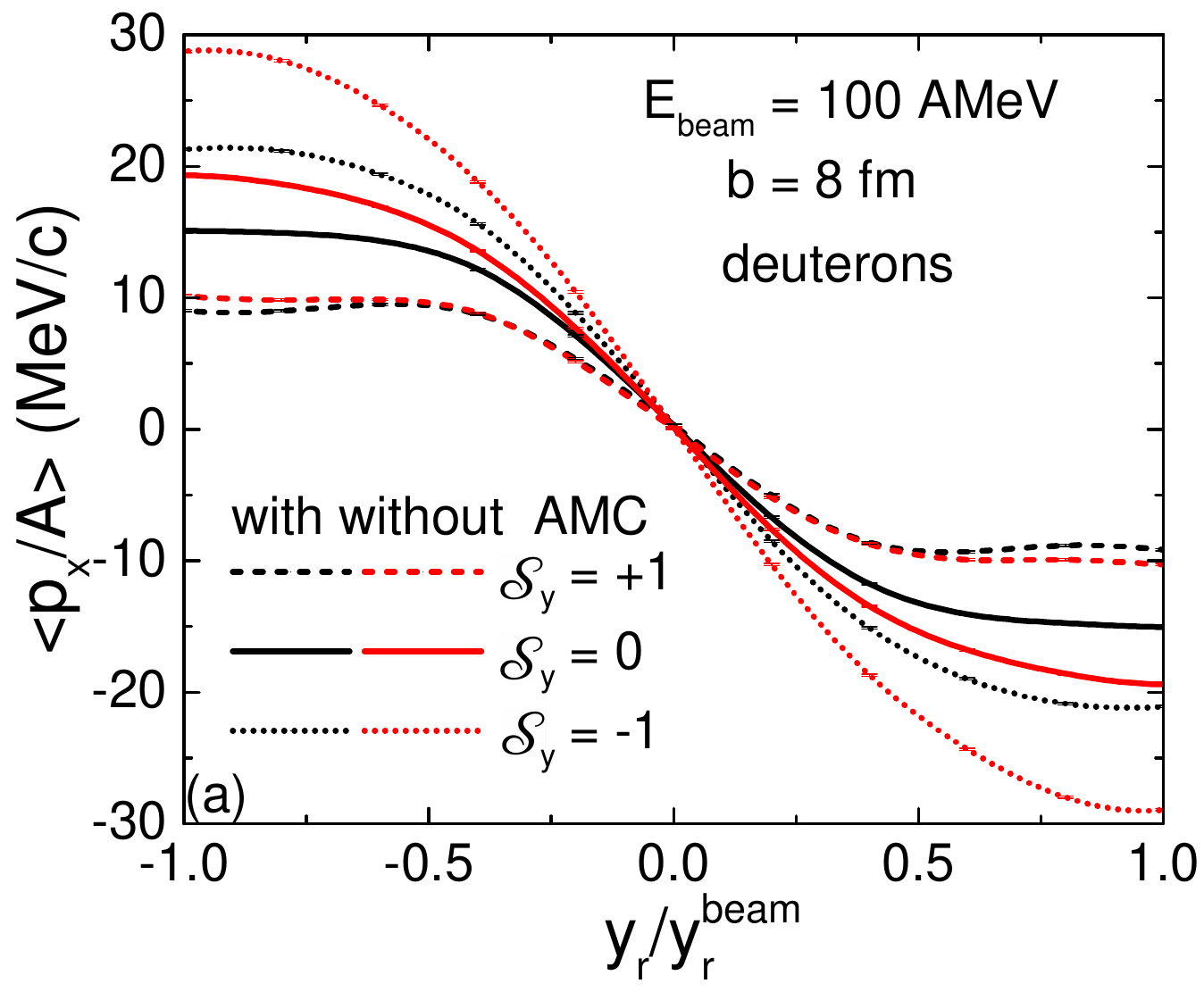}\includegraphics[width=0.373\linewidth]{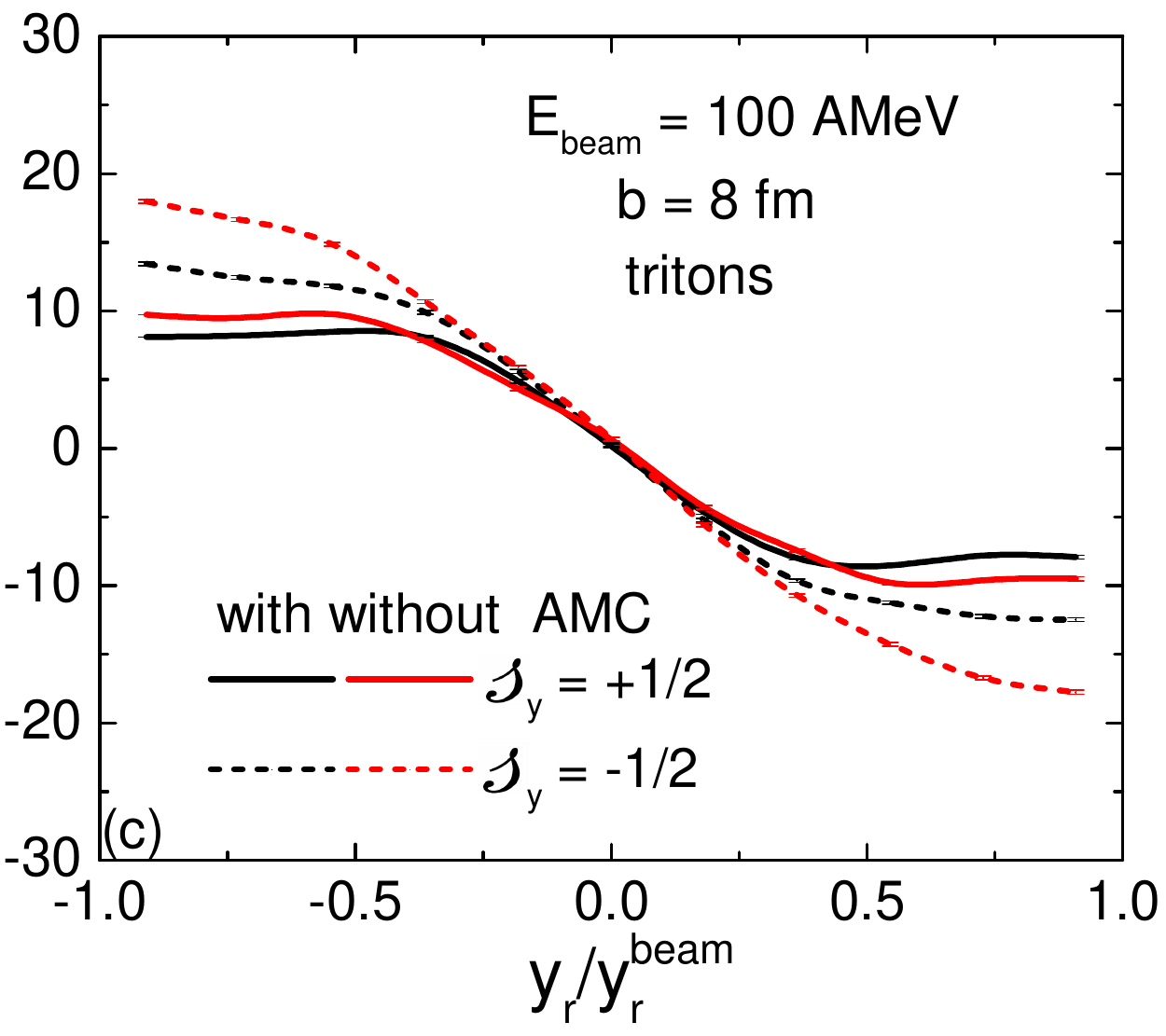}\\
\includegraphics[width=0.4\linewidth]{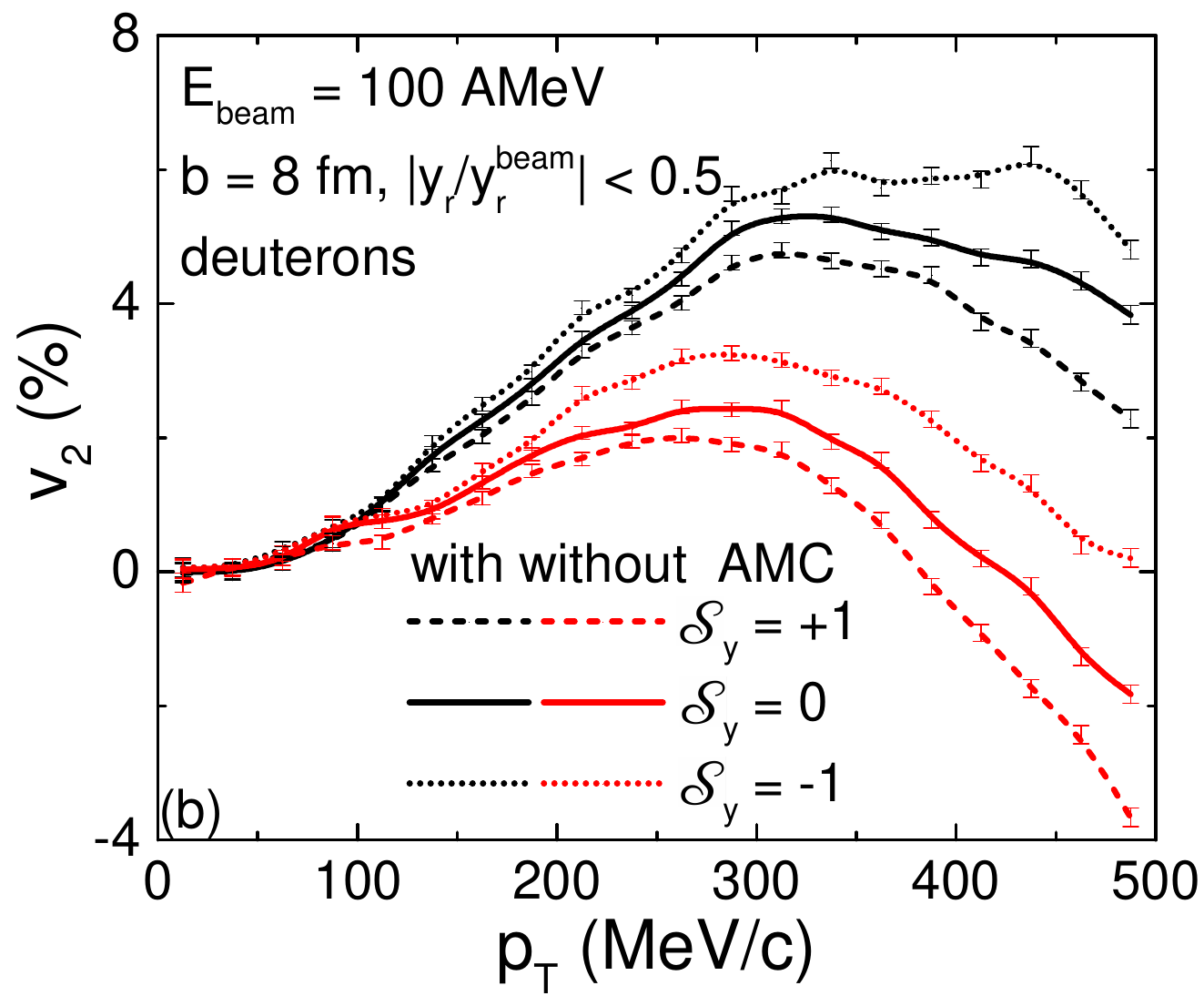}\includegraphics[width=0.373\linewidth]{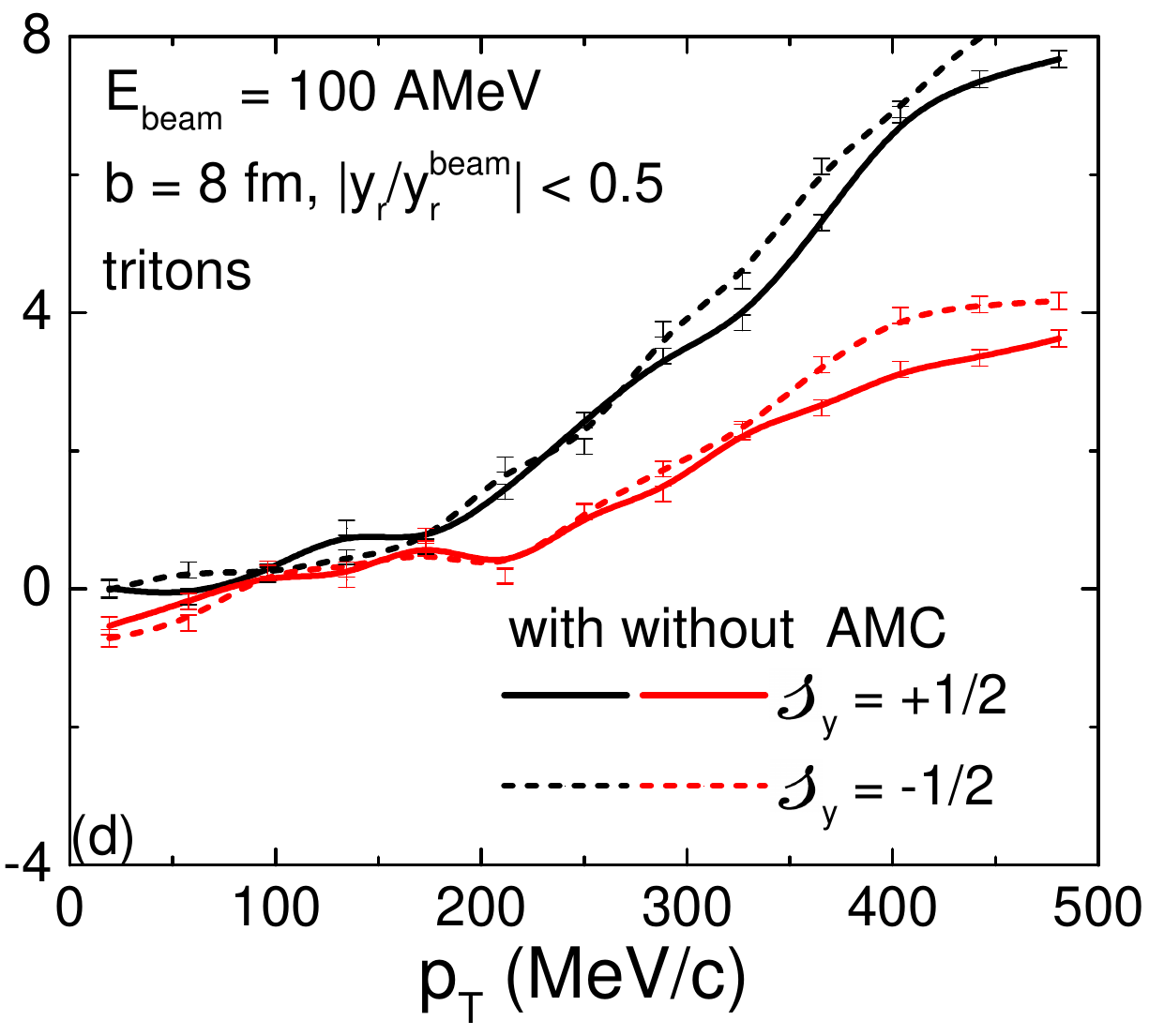}
\caption{\label{flowd} Transverse flow as a function of reduced rapidity $y_r/y_r^{beam}$ (upper) and elliptic flow as a function of transverse momentum $p_T$ (lower) for deuterons at spin states $\mathcal{S}_y=+1$, 0, and $-1$ (left) and for spin-up ($\mathcal{s}_y=+\frac{1}{2}$) and spin-down ($\mathcal{s}_y=-\frac{1}{2}$) tritons (right) in Au+Au collisions at the beam energy of 100 AMeV and impact parameter $\text{b}=8$ fm with and without angular momentum conservation. }
\end{figure*}

We now turn to the flows of nucleons in different spin states. Due to the coupling between the nucleon spin and the angular momentum in $y$ direction, nucleons with different spins $\mathcal{s}_y=\frac{1}{2}\sigma_y$ are expected to be affected by different spin-orbit potentials, and this is called the spin-Hall effect~\cite{PhysRevLett.83.1834}. As shown in the upper panel of Fig.~\ref{px}, nucleons with $\mathcal{s}_y=-\frac{1}{2} (+\frac{1}{2})$ are affected by a more repulsive (attractive) spin-orbit potential, and they thus have a stronger (weaker) transverse flow. We can define the spin up-down differential transverse flow as~\cite{Xu:2012hh}
\begin{equation}
F_{ud}(y_r) = \frac{1}{N(y_r)} \sum_{i=1}^{N(y_r)} (\sigma_y)_i (p_x)_i,
\end{equation}
where $N(y_r)$ is the nucleon number at rapidity $y_r$, and $(\sigma_y)_i$ and $(p_x)_i$ are the $y$ component of the spin and the momentum in $x$ direction for the $i$th nucleon. As shown in the middle panel of Fig.~\ref{px}, the spin up-down differential transverse flow has an appreciable magnitude, and it actually characterizes the strength of the spin-orbit potential~\cite{Xu:2012hh,Xia:2014rua}. The elliptic flow of free nucleons at midrapidities, as shown in the lower panel of Fig.~\ref{px}, also demonstrates the spin splitting, with nucleons of spin $\mathcal{s}_y=-\frac{1}{2} (+\frac{1}{2})$ affected by a more repulsive (attractive) spin-orbit potential and thus leading to a stronger (weaker) $v_2$. Incorporating the constraint of AMC, a lower density is reached, and this generally leads to a weaker transverse flow as shown in Fig.~\ref{px} (a), and also weakens slightly the spin up-down differential transverse flow in Fig.~\ref{px} (b). The preferred in-plane collisions with AMC generally enhances the in-plane flow and thus $v_2$ at lower $p_T$, but weakens the squeeze-out flow in $\pm y$ direction and thus leads to less negative $v_2$ at higher $p_T$.

The spin-dependent flows of free nucleons may also lead to the spin splitting of flows for light clusters at different spin states, and this is displayed in Fig.~\ref{flowd} where the productions of deuterons and tritons are calculated from the spin-dependent coalescence method. Note here the transverse flow is divided by the constituent nucleon number, i.e., 2 for deuterons and 3 for tritons. It is seen that the spin splittings of both transverse flows and elliptic flows are stronger for deuterons, compared with those for tritons. This is understandable, since deuterons at spin states $\mathcal{S}_y=\pm 1$ and 0 are formed by respectively the spin-triplet and spin-singlet state of neutron-proton pairs, so the spin effect are doubled or cancelled, while the spin effect for tritons at $\mathcal{s}_y=+\frac{1}{2}$ and $-\frac{1}{2}$ is determined by the residue neutron apart from the spin-singlet neutron-proton pair. The results are similar for $^3$He. Again, the constraint of AMC leads to systematically weaker transverse flows and stronger elliptic flows, at least for larger $p_T$ values.

\subsection{Spin-averaged flows}

\begin{figure}[ht]
\includegraphics[width=0.8\linewidth]{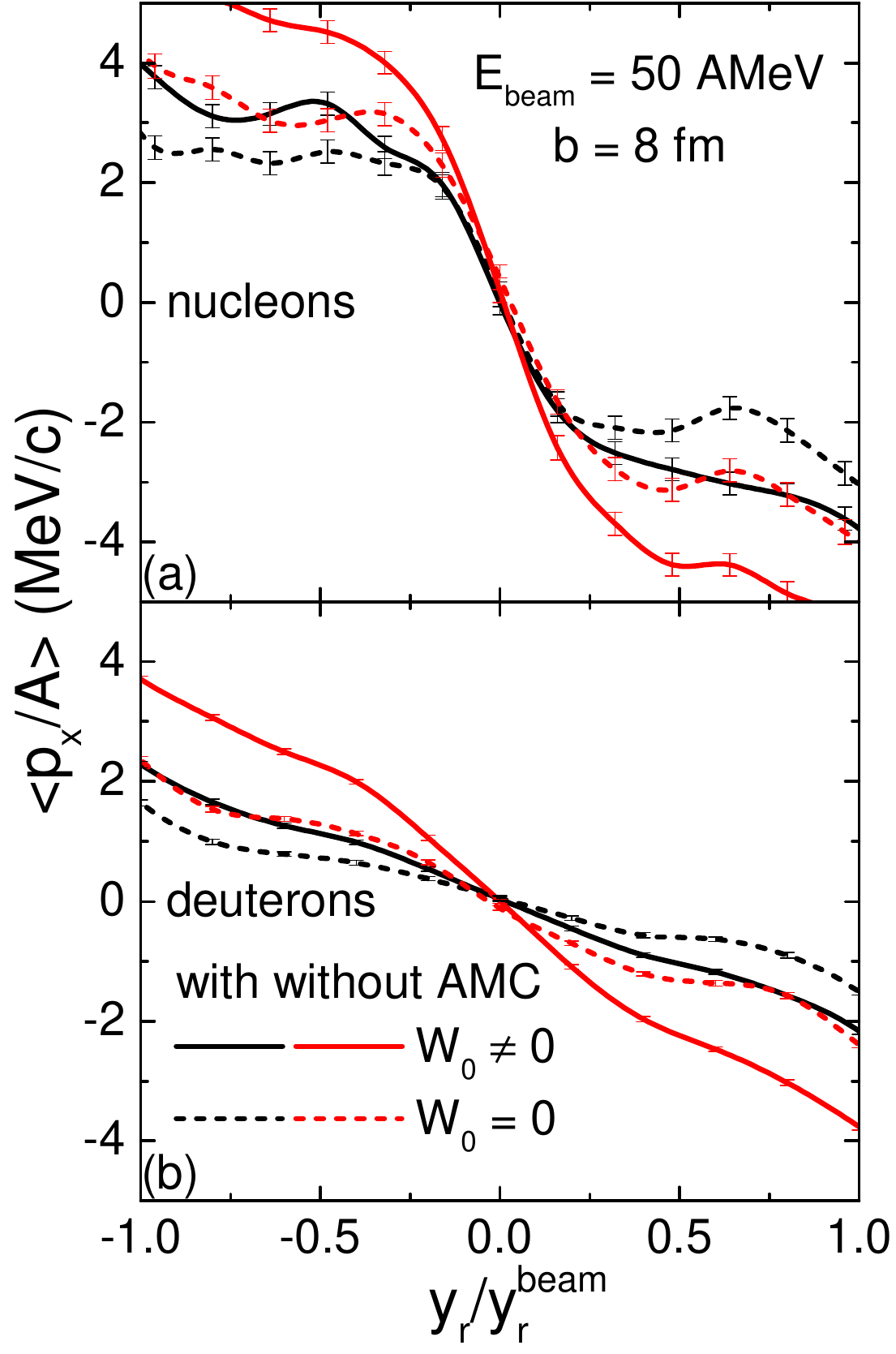}
\caption{\label{save} Spin-averaged transverse flows of free nucleons (a) and deuterons (b) as a function of reduced rapidity $y_r/y_r^{beam}$ in Au+Au collisions at the beam energy of 50 AMeV and impact parameter $\text{b}=8$ fm with and without angular momentum conservation and/or spin-dependent mean-field potential.}
\end{figure}

While the spin-dependent potential does leads to the spin polarization and the spin-dependent flows, one may argue that the spin-averaged observables predicted by traditional transport models are still valid. This is only partially true when the collision energy is relatively high. At lower collision energies, the dissipation effect from the spin-dependent potential is non-negligible. As seen in fusion reactions, the incorporation of the spin-dependent potential affects significantly the fusion threshold and the fusion cross section~\cite{PhysRevLett.56.2793,Maruhn:2006uh,Reinhard:1988zz}. Here we compare the spin-averaged transverse flow at $E_{lab}=50$ AMeV in Fig.~\ref{save}. With the spin-dependent potential ($W_0 \neq 0$), the stronger dissipation effect leads to a stronger transverse flow for free nucleons, compared to the case without the spin-dependent potential ($W_0=0$). The effect on the slope of the transverse flow is further enhanced when we consider deuterons from the coalescence of nucleons. Comparable to the effect of the spin-dependent potential, the incorporation of the AMC constraint leads to a systematical reduction of the transverse flow. It is thus seen that one should incorporate properly the spin-dependent mean-field potential and the constraint of AMC to predict accurately the transverse flow.

\section{Conclusion and outlook}
\label{summary}

We have improved the framework of the spin- and isospin-dependent Boltzmann-Uehling-Uhlenbeck (SIBUU) transport model, in the aspects of spin-dependent nucleon-nucleon collisions as well as the coalescence method, and particularly by incorporating the constraint of rigorous angular momentum conservation. We have revisited the global and local spin polarization in intermediate-energy heavy-ion collisions, as well as the spin-dependent transverse flows and elliptic flows, for spin-half nucleons, spin-one deuterons, and spin-half tritons/$^3$He. While the constraint of rigorous angular momentum conservation generally leads to weaker spin polarization and have some influence on the flows, the effects of the spin-orbit potential on the spin dynamics and spin-related observables are still appreciable. We have further demonstrated that the spin-orbit potential has a non-negligible effect on the spin-averaged transverse flows of nucleons and light clusters at low collision energies.

Experimentally, it is difficult to identify different spin states for nucleons or light clusters. On the other hand, the spin polarizations of hyperons or the spin alignment of vector mesons are measurable through the angular distribution of their decays. It is thus of great interest to incorporate the relevant inelastic collision channels for the production of these particles, and see whether their spins will be affected by the nucleon spin dynamics. The well-developed SIBUU model in the present study could then be used to study of spin dynamics in heavy-ion collisions at the beam energy of a few AGeV, or even used as a hadronic afterburner for the spin-relevant studies in relativistic heavy-ion collisions.

\begin{acknowledgments}
This work is supported by the Strategic Priority Research Program of the Chinese Academy of Sciences under Grant No. XDB34030000, the National Natural Science Foundation of China under Grant Nos. 12375125, 11922514, and 11475243, and the Fundamental Research Funds for the Central Universities.
\end{acknowledgments}

\appendix

\section{Definition of spin states}
\label{app}

The expectation value of a nucleon spin is determined by the azimuthal angle $\phi$ and the polar angle $\theta$
\begin{equation} \label{epspin}
\vec{\sigma}=(\sin\theta\cos\phi,\sin\theta\sin\phi,\cos\theta),
\end{equation}
and its spin state can be written as
\begin{equation}
\chi=\begin{pmatrix}e^{\frac{-i\phi}{2}}\cos{\frac{\theta}{2}}
\\e^{\frac{i\phi}{2}}\sin{\frac{\theta}{2}}
\end{pmatrix}.
\end{equation}
The expectation values of the spin in $x$, $y$, and $z$ directions are
\begin{eqnarray}
\overline{\sigma_{x}}&=&\chi^+\sigma_x\chi=\sin\theta\cos\phi,\\
\overline{\sigma_{y}}&=&\chi^+\sigma_y\chi=\sin\theta\sin\phi,\\
\overline{\sigma_{z}}&=&\chi^+\sigma_z\chi=\cos\theta.
\end{eqnarray}
We can define the two-nucleon spin state as the direct product of the spin state of each nucleon
\begin{equation}\label{phi}
\Psi=\chi_1\otimes \chi_2=\begin{pmatrix}
e^{\frac{-i(\phi_1+\phi_2)}{2}}\cos{\frac{\theta_1}{2}}\cos{\frac{\theta_2}{2}}\\
e^{\frac{-i(\phi_1-\phi_2)}{2}}\cos{\frac{\theta_1}{2}}\sin{\frac{\theta_2}{2}}\\
e^{\frac{i(\phi_1-\phi_2)}{2}}\sin{\frac{\theta_1}{2}}\cos{\frac{\theta_2}{2}}\\
e^{\frac{i(\phi_1+\phi_2)}{2}}\sin{\frac{\theta_1}{2}}\sin{\frac{\theta_2}{2}}
\end{pmatrix}.
\end{equation}
In this way, the expectation values of the spin in the $a=x$, $y$, and $z$ directions for nucleon 1 and nucleon 2 can be expressed as
\begin{eqnarray}
(\overline{\sigma_a})_1&=&\chi_1^+\sigma_a\chi_1=\Psi^+(\sigma_a)_1\Psi,\\
(\overline{\sigma_a})_2&=&\chi_2^+\sigma_a\chi_2=\Psi^+(\sigma_a)_2\Psi,
\end{eqnarray}
with
\begin{eqnarray}
(\sigma_a)_1&=&\sigma_a\otimes I_{2\times2},\\
(\sigma_a)_2&=&I_{2\times2}\otimes\sigma_a.
\end{eqnarray}
With the spin-up and spin-down states for each nucleon pair defined as
\begin{eqnarray}
\left|\uparrow\uparrow\right\rangle =\begin{pmatrix}1\\0\\0\\0\end{pmatrix},~
\left|\uparrow\downarrow\right\rangle =\begin{pmatrix}0\\1\\0\\0\end{pmatrix},~
\left|\downarrow\uparrow\right\rangle =\begin{pmatrix}0\\0\\1\\0\end{pmatrix},~
\left|\downarrow\downarrow\right\rangle =\begin{pmatrix}0\\0\\0\\1\end{pmatrix}, \notag
\end{eqnarray}
the spin-singlet and the spin-triplet states of two nucleons can be expressed respectively as
\begin{eqnarray}\label{singlet}
\chi_{0,0}=\begin{pmatrix}0\\\frac{1}{\sqrt2}\\-\frac{1}{\sqrt2}\\0\end{pmatrix}
\end{eqnarray}
and
\begin{eqnarray}\label{triplet}
\chi_{1,1}=\begin{pmatrix}1\\0\\0\\0\end{pmatrix},~
\chi_{1,0}=\begin{pmatrix}0\\\frac{1}{\sqrt2}\\\frac{1}{\sqrt2}\\0\end{pmatrix},~
\chi_{1,-1}=\begin{pmatrix}0\\0\\0\\1\end{pmatrix}.
\end{eqnarray}
By taking the product of Eq.~(\ref{phi}) with Eqs.~(\ref{singlet}) and (\ref{triplet}), we can calculate the probability of a nucleon pair in the spin-singlet state and in the spin-triplet state, and the normalization condition
\begin{equation}
\left| \left< \chi _{0,0} \mid \Psi \right> \right|^2 + \left| \left< \chi _{1,1} \mid \Psi \right> \right|^2 + \left| \left< \chi _{1,0} \mid \Psi \right> \right|^2 +\left| \left< \chi _{1,-1} \mid \Psi \right> \right|^2 = 1 \notag
\end{equation}
is satisfied. The above definition of the spin state is general, since we can rotate the polar angle $\theta$ and the azimuthal angle $\phi$ in Eq.~(\ref{epspin}) so that the projection of the spin can be performed in an arbitrary direction.
\bibliography{sibuu}

\begin{thebibliography}{57}%
\makeatletter
\providecommand \@ifxundefined [1]{%
 \@ifx{#1\undefined}
}%
\providecommand \@ifnum [1]{%
 \ifnum #1\expandafter \@firstoftwo
 \else \expandafter \@secondoftwo
 \fi
}%
\providecommand \@ifx [1]{%
 \ifx #1\expandafter \@firstoftwo
 \else \expandafter \@secondoftwo
 \fi
}%
\providecommand \natexlab [1]{#1}%
\providecommand \enquote  [1]{``#1''}%
\providecommand \bibnamefont  [1]{#1}%
\providecommand \bibfnamefont [1]{#1}%
\providecommand \citenamefont [1]{#1}%
\providecommand \href@noop [0]{\@secondoftwo}%
\providecommand \href [0]{\begingroup \@sanitize@url \@href}%
\providecommand \@href[1]{\@@startlink{#1}\@@href}%
\providecommand \@@href[1]{\endgroup#1\@@endlink}%
\providecommand \@sanitize@url [0]{\catcode `\\12\catcode `\$12\catcode
  `\&12\catcode `\#12\catcode `\^12\catcode `\_12\catcode `\%12\relax}%
\providecommand \@@startlink[1]{}%
\providecommand \@@endlink[0]{}%
\providecommand \url  [0]{\begingroup\@sanitize@url \@url }%
\providecommand \@url [1]{\endgroup\@href {#1}{\urlprefix }}%
\providecommand \urlprefix  [0]{URL }%
\providecommand \Eprint [0]{\href }%
\providecommand \doibase [0]{http://dx.doi.org/}%
\providecommand \selectlanguage [0]{\@gobble}%
\providecommand \bibinfo  [0]{\@secondoftwo}%
\providecommand \bibfield  [0]{\@secondoftwo}%
\providecommand \translation [1]{[#1]}%
\providecommand \BibitemOpen [0]{}%
\providecommand \bibitemStop [0]{}%
\providecommand \bibitemNoStop [0]{.\EOS\space}%
\providecommand \EOS [0]{\spacefactor3000\relax}%
\providecommand \BibitemShut  [1]{\csname bibitem#1\endcsname}%
\let\auto@bib@innerbib\@empty
\bibitem [{\citenamefont {Kharzeev}\ \emph {et~al.}(2016)\citenamefont
  {Kharzeev}, \citenamefont {Liao}, \citenamefont {Voloshin},\ and\
  \citenamefont {Wang}}]{Kharzeev:2015znc}%
  \BibitemOpen
  \bibfield  {author} {\bibinfo {author} {\bibfnamefont {D.~E.}\ \bibnamefont
  {Kharzeev}}, \bibinfo {author} {\bibfnamefont {J.}~\bibnamefont {Liao}},
  \bibinfo {author} {\bibfnamefont {S.~A.}\ \bibnamefont {Voloshin}}, \ and\
  \bibinfo {author} {\bibfnamefont {G.}~\bibnamefont {Wang}},\ }\bibfield
  {title} {\enquote {\bibinfo {title} {{Chiral magnetic and vortical effects in
  high-energy nuclear collisions\textemdash{}A status report}},}\ }\href
  {\doibase 10.1016/j.ppnp.2016.01.001} {\bibfield  {journal} {\bibinfo
  {journal} {Prog. Part. Nucl. Phys.}\ }\textbf {\bibinfo {volume} {88}},\
  \bibinfo {pages} {1} (\bibinfo {year} {2016})}\BibitemShut {NoStop}%
\bibitem [{\citenamefont {Huang}(2016)}]{Huang:2015oca}%
  \BibitemOpen
  \bibfield  {author} {\bibinfo {author} {\bibfnamefont {X.~G.}\ \bibnamefont
  {Huang}},\ }\bibfield  {title} {\enquote {\bibinfo {title} {{Electromagnetic
  fields and anomalous transports in heavy-ion collisions --- A pedagogical
  review}},}\ }\href {\doibase 10.1088/0034-4885/79/7/076302} {\bibfield
  {journal} {\bibinfo  {journal} {Rept. Prog. Phys.}\ }\textbf {\bibinfo
  {volume} {79}},\ \bibinfo {pages} {076302} (\bibinfo {year}
  {2016})}\BibitemShut {NoStop}%
\bibitem [{\citenamefont {Becattini}\ and\ \citenamefont
  {Lisa}(2020)}]{Becattini:2020ngo}%
  \BibitemOpen
  \bibfield  {author} {\bibinfo {author} {\bibfnamefont {F.}~\bibnamefont
  {Becattini}}\ and\ \bibinfo {author} {\bibfnamefont {M.~A.}\ \bibnamefont
  {Lisa}},\ }\bibfield  {title} {\enquote {\bibinfo {title} {{Polarization and
  Vorticity in the Quark\textendash{}Gluon Plasma}},}\ }\href {\doibase
  10.1146/annurev-nucl-021920-095245} {\bibfield  {journal} {\bibinfo
  {journal} {Ann. Rev. Nucl. Part. Sci.}\ }\textbf {\bibinfo {volume} {70}},\
  \bibinfo {pages} {395} (\bibinfo {year} {2020})}\BibitemShut {NoStop}%
\bibitem [{\citenamefont {Fukushima}\ \emph {et~al.}(2008)\citenamefont
  {Fukushima}, \citenamefont {Kharzeev},\ and\ \citenamefont
  {Warringa}}]{Fukushima:2008xe}%
  \BibitemOpen
  \bibfield  {author} {\bibinfo {author} {\bibfnamefont {K.}~\bibnamefont
  {Fukushima}}, \bibinfo {author} {\bibfnamefont {D.~E.}\ \bibnamefont
  {Kharzeev}}, \ and\ \bibinfo {author} {\bibfnamefont {H.~J.}\ \bibnamefont
  {Warringa}},\ }\bibfield  {title} {\enquote {\bibinfo {title} {{The Chiral
  Magnetic Effect}},}\ }\href {\doibase 10.1103/PhysRevD.78.074033} {\bibfield
  {journal} {\bibinfo  {journal} {Phys. Rev. D}\ }\textbf {\bibinfo {volume}
  {78}},\ \bibinfo {pages} {074033} (\bibinfo {year} {2008})}\BibitemShut
  {NoStop}%
\bibitem [{\citenamefont {Burnier}\ \emph {et~al.}(2011)\citenamefont
  {Burnier}, \citenamefont {Kharzeev}, \citenamefont {Liao},\ and\
  \citenamefont {Yee}}]{Burnier:2011bf}%
  \BibitemOpen
  \bibfield  {author} {\bibinfo {author} {\bibfnamefont {Y.}~\bibnamefont
  {Burnier}}, \bibinfo {author} {\bibfnamefont {D.~E.}\ \bibnamefont
  {Kharzeev}}, \bibinfo {author} {\bibfnamefont {J.}~\bibnamefont {Liao}}, \
  and\ \bibinfo {author} {\bibfnamefont {H.~U.}\ \bibnamefont {Yee}},\
  }\bibfield  {title} {\enquote {\bibinfo {title} {{Chiral magnetic wave at
  finite baryon density and the electric quadrupole moment of quark-gluon
  plasma in heavy ion collisions}},}\ }\href {\doibase
  10.1103/PhysRevLett.107.052303} {\bibfield  {journal} {\bibinfo  {journal}
  {Phys. Rev. Lett.}\ }\textbf {\bibinfo {volume} {107}},\ \bibinfo {pages}
  {052303} (\bibinfo {year} {2011})}\BibitemShut {NoStop}%
\bibitem [{\citenamefont {Kharzeev}\ and\ \citenamefont
  {Zhitnitsky}(2007)}]{Kharzeev:2007tn}%
  \BibitemOpen
  \bibfield  {author} {\bibinfo {author} {\bibfnamefont {D.}~\bibnamefont
  {Kharzeev}}\ and\ \bibinfo {author} {\bibfnamefont {A.}~\bibnamefont
  {Zhitnitsky}},\ }\bibfield  {title} {\enquote {\bibinfo {title} {{Charge
  separation induced by P-odd bubbles in QCD matter}},}\ }\href {\doibase
  10.1016/j.nuclphysa.2007.10.001} {\bibfield  {journal} {\bibinfo  {journal}
  {Nucl. Phys. A}\ }\textbf {\bibinfo {volume} {797}},\ \bibinfo {pages} {67}
  (\bibinfo {year} {2007})}\BibitemShut {NoStop}%
\bibitem [{\citenamefont {Abelev}\ \emph {et~al.}(2009)\citenamefont {Abelev}
  \emph {et~al.}}]{STAR:2009wot}%
  \BibitemOpen
  \bibfield  {author} {\bibinfo {author} {\bibfnamefont {B.~I.}\ \bibnamefont
  {Abelev}} \emph {et~al.} (\bibinfo {collaboration} {STAR}),\ }\bibfield
  {title} {\enquote {\bibinfo {title} {{Azimuthal Charged-Particle Correlations
  and Possible Local Strong Parity Violation}},}\ }\href {\doibase
  10.1103/PhysRevLett.103.251601} {\bibfield  {journal} {\bibinfo  {journal}
  {Phys. Rev. Lett.}\ }\textbf {\bibinfo {volume} {103}},\ \bibinfo {pages}
  {251601} (\bibinfo {year} {2009})}\BibitemShut {NoStop}%
\bibitem [{\citenamefont {Abelev}\ \emph {et~al.}(2013)\citenamefont {Abelev}
  \emph {et~al.}}]{ALICE:2012nhw}%
  \BibitemOpen
  \bibfield  {author} {\bibinfo {author} {\bibfnamefont {B.}~\bibnamefont
  {Abelev}} \emph {et~al.} (\bibinfo {collaboration} {ALICE}),\ }\bibfield
  {title} {\enquote {\bibinfo {title} {{Charge separation relative to the
  reaction plane in Pb-Pb collisions at $\sqrt{s_{NN}}= 2.76$ TeV}},}\ }\href
  {\doibase 10.1103/PhysRevLett.110.012301} {\bibfield  {journal} {\bibinfo
  {journal} {Phys. Rev. Lett.}\ }\textbf {\bibinfo {volume} {110}},\ \bibinfo
  {pages} {012301} (\bibinfo {year} {2013})}\BibitemShut {NoStop}%
\bibitem [{\citenamefont {Adamczyk}\ \emph {et~al.}(2015)\citenamefont
  {Adamczyk} \emph {et~al.}}]{STAR:2015wza}%
  \BibitemOpen
  \bibfield  {author} {\bibinfo {author} {\bibfnamefont {L.}~\bibnamefont
  {Adamczyk}} \emph {et~al.} (\bibinfo {collaboration} {STAR}),\ }\bibfield
  {title} {\enquote {\bibinfo {title} {{Observation of charge asymmetry
  dependence of pion elliptic flow and the possible chiral magnetic wave in
  heavy-ion collisions}},}\ }\href {\doibase 10.1103/PhysRevLett.114.252302}
  {\bibfield  {journal} {\bibinfo  {journal} {Phys. Rev. Lett.}\ }\textbf
  {\bibinfo {volume} {114}},\ \bibinfo {pages} {252302} (\bibinfo {year}
  {2015})}\BibitemShut {NoStop}%
\bibitem [{\citenamefont {Zhao}(2014)}]{Zhao:2014aja}%
  \BibitemOpen
  \bibfield  {author} {\bibinfo {author} {\bibfnamefont {F.}~\bibnamefont
  {Zhao}} (\bibinfo {collaboration} {STAR}),\ }\bibfield  {title} {\enquote
  {\bibinfo {title} {{$\Lambda(K_S^0)-h^\pm$ and $\Lambda-p$ azimuthal
  correlations with respect to event plane and search for chiral magnetic and
  vortical effects}},}\ }\href {\doibase 10.1016/j.nuclphysa.2014.08.108}
  {\bibfield  {journal} {\bibinfo  {journal} {Nucl. Phys. A}\ }\textbf
  {\bibinfo {volume} {931}},\ \bibinfo {pages} {746} (\bibinfo {year}
  {2014})}\BibitemShut {NoStop}%
\bibitem [{\citenamefont {Abdallah}\ \emph {et~al.}(2023)\citenamefont
  {Abdallah} \emph {et~al.}}]{STAR:2022fan}%
  \BibitemOpen
  \bibfield  {author} {\bibinfo {author} {\bibfnamefont {M.~S.}\ \bibnamefont
  {Abdallah}} \emph {et~al.} (\bibinfo {collaboration} {STAR}),\ }\bibfield
  {title} {\enquote {\bibinfo {title} {{Pattern of global spin alignment of
  \ensuremath{\phi} and K$^{*0}$ mesons in heavy-ion collisions}},}\ }\href
  {\doibase 10.1038/s41586-022-05557-5} {\bibfield  {journal} {\bibinfo
  {journal} {Nature}\ }\textbf {\bibinfo {volume} {614}},\ \bibinfo {pages}
  {244} (\bibinfo {year} {2023})}\BibitemShut {NoStop}%
\bibitem [{\citenamefont {Becattini}\ \emph {et~al.}(2019)\citenamefont
  {Becattini}, \citenamefont {Cao},\ and\ \citenamefont
  {Speranza}}]{Becattini:2019ntv}%
  \BibitemOpen
  \bibfield  {author} {\bibinfo {author} {\bibfnamefont {F.}~\bibnamefont
  {Becattini}}, \bibinfo {author} {\bibfnamefont {G.}~\bibnamefont {Cao}}, \
  and\ \bibinfo {author} {\bibfnamefont {E.}~\bibnamefont {Speranza}},\
  }\bibfield  {title} {\enquote {\bibinfo {title} {{Polarization transfer in
  hyperon decays and its effect in relativistic nuclear collisions}},}\ }\href
  {\doibase 10.1140/epjc/s10052-019-7213-6} {\bibfield  {journal} {\bibinfo
  {journal} {Eur. Phys. J. C}\ }\textbf {\bibinfo {volume} {79}},\ \bibinfo
  {pages} {741} (\bibinfo {year} {2019})}\BibitemShut {NoStop}%
\bibitem [{\citenamefont {Xia}\ \emph {et~al.}(2019)\citenamefont {Xia},
  \citenamefont {Li}, \citenamefont {Huang},\ and\ \citenamefont
  {Huang}}]{Xia:2019fjf}%
  \BibitemOpen
  \bibfield  {author} {\bibinfo {author} {\bibfnamefont {X.~L.}\ \bibnamefont
  {Xia}}, \bibinfo {author} {\bibfnamefont {H.}~\bibnamefont {Li}}, \bibinfo
  {author} {\bibfnamefont {X.~G.}\ \bibnamefont {Huang}}, \ and\ \bibinfo
  {author} {\bibfnamefont {H.~Z.}\ \bibnamefont {Huang}},\ }\bibfield  {title}
  {\enquote {\bibinfo {title} {{Feed-down effect on \ensuremath{\Lambda} spin
  polarization}},}\ }\href {\doibase 10.1103/PhysRevC.100.014913} {\bibfield
  {journal} {\bibinfo  {journal} {Phys. Rev. C}\ }\textbf {\bibinfo {volume}
  {100}},\ \bibinfo {pages} {014913} (\bibinfo {year} {2019})}\BibitemShut
  {NoStop}%
\bibitem [{\citenamefont {Xu}\ \emph {et~al.}(2015)\citenamefont {Xu},
  \citenamefont {Li}, \citenamefont {Shen},\ and\ \citenamefont
  {Xia}}]{Xu:2015kxa}%
  \BibitemOpen
  \bibfield  {author} {\bibinfo {author} {\bibfnamefont {J.}~\bibnamefont
  {Xu}}, \bibinfo {author} {\bibfnamefont {B.~A.}\ \bibnamefont {Li}}, \bibinfo
  {author} {\bibfnamefont {W.~Q.}\ \bibnamefont {Shen}}, \ and\ \bibinfo
  {author} {\bibfnamefont {Y.}~\bibnamefont {Xia}},\ }\bibfield  {title}
  {\enquote {\bibinfo {title} {{Dynamical effects of spin-dependent
  interactions in low- and intermediate-energy heavy-ion reactions}},}\ }\href
  {\doibase 10.1007/s11467-015-0479-8} {\bibfield  {journal} {\bibinfo
  {journal} {Front. Phys. (Beijing)}\ }\textbf {\bibinfo {volume} {10}},\
  \bibinfo {pages} {102501} (\bibinfo {year} {2015})}\BibitemShut {NoStop}%
\bibitem [{\citenamefont {Mayer}(1948)}]{Mayer:1948zz}%
  \BibitemOpen
  \bibfield  {author} {\bibinfo {author} {\bibfnamefont {M.~G.}\ \bibnamefont
  {Mayer}},\ }\bibfield  {title} {\enquote {\bibinfo {title} {{On Closed Shells
  in Nuclei}},}\ }\href {\doibase 10.1103/PhysRev.74.235} {\bibfield  {journal}
  {\bibinfo  {journal} {Phys. Rev.}\ }\textbf {\bibinfo {volume} {74}},\
  \bibinfo {pages} {235} (\bibinfo {year} {1948})}\BibitemShut {NoStop}%
\bibitem [{\citenamefont {Mayer}(1949)}]{Mayer:1949pd}%
  \BibitemOpen
  \bibfield  {author} {\bibinfo {author} {\bibfnamefont {M.~G.}\ \bibnamefont
  {Mayer}},\ }\bibfield  {title} {\enquote {\bibinfo {title} {{On Closed Shells
  in Nuclei. 2}},}\ }\href {\doibase 10.1103/PhysRev.75.1969} {\bibfield
  {journal} {\bibinfo  {journal} {Phys. Rev.}\ }\textbf {\bibinfo {volume}
  {75}},\ \bibinfo {pages} {1969} (\bibinfo {year} {1949})}\BibitemShut
  {NoStop}%
\bibitem [{\citenamefont {Haxel}\ \emph {et~al.}(1949)\citenamefont {Haxel},
  \citenamefont {Jensen},\ and\ \citenamefont {Suess}}]{Haxel:1949fjd}%
  \BibitemOpen
  \bibfield  {author} {\bibinfo {author} {\bibfnamefont {O.}~\bibnamefont
  {Haxel}}, \bibinfo {author} {\bibfnamefont {J.~H.~D.}\ \bibnamefont
  {Jensen}}, \ and\ \bibinfo {author} {\bibfnamefont {H.~E.}\ \bibnamefont
  {Suess}},\ }\bibfield  {title} {\enquote {\bibinfo {title} {{On the ``Magic
  Numbers'' in Nuclear Structure}},}\ }\href {\doibase
  10.1103/PhysRev.75.1766.2} {\bibfield  {journal} {\bibinfo  {journal} {Phys.
  Rev.}\ }\textbf {\bibinfo {volume} {75}},\ \bibinfo {pages} {1766} (\bibinfo
  {year} {1949})}\BibitemShut {NoStop}%
\bibitem [{\citenamefont {Otsuka}\ \emph {et~al.}(2005)\citenamefont {Otsuka},
  \citenamefont {Suzuki}, \citenamefont {Fujimoto}, \citenamefont {Grawe},\
  and\ \citenamefont {Akaishi}}]{Otsuka:2005zz}%
  \BibitemOpen
  \bibfield  {author} {\bibinfo {author} {\bibfnamefont {T.}~\bibnamefont
  {Otsuka}}, \bibinfo {author} {\bibfnamefont {T.}~\bibnamefont {Suzuki}},
  \bibinfo {author} {\bibfnamefont {R.}~\bibnamefont {Fujimoto}}, \bibinfo
  {author} {\bibfnamefont {H.}~\bibnamefont {Grawe}}, \ and\ \bibinfo {author}
  {\bibfnamefont {Y.}~\bibnamefont {Akaishi}},\ }\bibfield  {title} {\enquote
  {\bibinfo {title} {{Evolution of Nuclear Shells due to the Tensor Force}},}\
  }\href {\doibase 10.1103/PhysRevLett.95.232502} {\bibfield  {journal}
  {\bibinfo  {journal} {Phys. Rev. Lett.}\ }\textbf {\bibinfo {volume} {95}},\
  \bibinfo {pages} {232502} (\bibinfo {year} {2005})}\BibitemShut {NoStop}%
\bibitem [{\citenamefont {Otsuka}\ \emph {et~al.}(2006)\citenamefont {Otsuka},
  \citenamefont {Matsuo},\ and\ \citenamefont {Abe}}]{Otsuka:2006zz}%
  \BibitemOpen
  \bibfield  {author} {\bibinfo {author} {\bibfnamefont {T.}~\bibnamefont
  {Otsuka}}, \bibinfo {author} {\bibfnamefont {T.}~\bibnamefont {Matsuo}}, \
  and\ \bibinfo {author} {\bibfnamefont {D.}~\bibnamefont {Abe}},\ }\bibfield
  {title} {\enquote {\bibinfo {title} {{Mean Field with Tensor Force and Shell
  Structure of Exotic Nuclei}},}\ }\href {\doibase
  10.1103/PhysRevLett.97.162501} {\bibfield  {journal} {\bibinfo  {journal}
  {Phys. Rev. Lett.}\ }\textbf {\bibinfo {volume} {97}},\ \bibinfo {pages}
  {162501} (\bibinfo {year} {2006})}\BibitemShut {NoStop}%
\bibitem [{\citenamefont {Otsuka}\ \emph {et~al.}(2010)\citenamefont {Otsuka},
  \citenamefont {Suzuki}, \citenamefont {Honma}, \citenamefont {Utsuno},
  \citenamefont {Tsunoda}, \citenamefont {Tsukiyama},\ and\ \citenamefont
  {Hjorth-Jensen}}]{Otsuka:2009qs}%
  \BibitemOpen
  \bibfield  {author} {\bibinfo {author} {\bibfnamefont {T.}~\bibnamefont
  {Otsuka}}, \bibinfo {author} {\bibfnamefont {T.}~\bibnamefont {Suzuki}},
  \bibinfo {author} {\bibfnamefont {M.}~\bibnamefont {Honma}}, \bibinfo
  {author} {\bibfnamefont {Y.}~\bibnamefont {Utsuno}}, \bibinfo {author}
  {\bibfnamefont {N.}~\bibnamefont {Tsunoda}}, \bibinfo {author} {\bibfnamefont
  {K.}~\bibnamefont {Tsukiyama}}, \ and\ \bibinfo {author} {\bibfnamefont
  {M.}~\bibnamefont {Hjorth-Jensen}},\ }\bibfield  {title} {\enquote {\bibinfo
  {title} {{Novel features of nuclear forces and shell evolution in exotic
  nuclei}},}\ }\href {\doibase 10.1103/PhysRevLett.104.012501} {\bibfield
  {journal} {\bibinfo  {journal} {Phys. Rev. Lett.}\ }\textbf {\bibinfo
  {volume} {104}},\ \bibinfo {pages} {012501} (\bibinfo {year}
  {2010})}\BibitemShut {NoStop}%
\bibitem [{\citenamefont {Umar}\ \emph {et~al.}(1986)\citenamefont {Umar},
  \citenamefont {Strayer},\ and\ \citenamefont
  {Reinhard}}]{PhysRevLett.56.2793}%
  \BibitemOpen
  \bibfield  {author} {\bibinfo {author} {\bibfnamefont {A.~S.}\ \bibnamefont
  {Umar}}, \bibinfo {author} {\bibfnamefont {M.~R.}\ \bibnamefont {Strayer}}, \
  and\ \bibinfo {author} {\bibfnamefont {P.~G.}\ \bibnamefont {Reinhard}},\
  }\bibfield  {title} {\enquote {\bibinfo {title} {Resolution of the fusion
  window anomaly in heavy-ion collisions},}\ }\href {\doibase
  10.1103/PhysRevLett.56.2793} {\bibfield  {journal} {\bibinfo  {journal}
  {Phys. Rev. Lett.}\ }\textbf {\bibinfo {volume} {56}},\ \bibinfo {pages}
  {2793} (\bibinfo {year} {1986})}\BibitemShut {NoStop}%
\bibitem [{\citenamefont {Maruhn}\ \emph {et~al.}(2006)\citenamefont {Maruhn},
  \citenamefont {Reinhard}, \citenamefont {Stevenson},\ and\ \citenamefont
  {Strayer}}]{Maruhn:2006uh}%
  \BibitemOpen
  \bibfield  {author} {\bibinfo {author} {\bibfnamefont {J.~A.}\ \bibnamefont
  {Maruhn}}, \bibinfo {author} {\bibfnamefont {P.~G.}\ \bibnamefont
  {Reinhard}}, \bibinfo {author} {\bibfnamefont {P.~D.}\ \bibnamefont
  {Stevenson}}, \ and\ \bibinfo {author} {\bibfnamefont {M.~R.}\ \bibnamefont
  {Strayer}},\ }\bibfield  {title} {\enquote {\bibinfo {title}
  {{Spin-excitation mechanisms in Skyrme-force time-dependent Hartree-Fock}},}\
  }\href {\doibase 10.1103/PhysRevC.74.027601} {\bibfield  {journal} {\bibinfo
  {journal} {Phys. Rev. C}\ }\textbf {\bibinfo {volume} {74}},\ \bibinfo
  {pages} {027601} (\bibinfo {year} {2006})}\BibitemShut {NoStop}%
\bibitem [{\citenamefont {Reinhard}\ \emph {et~al.}(1988)\citenamefont
  {Reinhard}, \citenamefont {Umar}, \citenamefont {Davies}, \citenamefont
  {Strayer},\ and\ \citenamefont {Lee}}]{Reinhard:1988zz}%
  \BibitemOpen
  \bibfield  {author} {\bibinfo {author} {\bibfnamefont {P.~G.}\ \bibnamefont
  {Reinhard}}, \bibinfo {author} {\bibfnamefont {A.~S.}\ \bibnamefont {Umar}},
  \bibinfo {author} {\bibfnamefont {K.~Thomas~R.}\ \bibnamefont {Davies}},
  \bibinfo {author} {\bibfnamefont {M.~R.}\ \bibnamefont {Strayer}}, \ and\
  \bibinfo {author} {\bibfnamefont {S.~J.}\ \bibnamefont {Lee}},\ }\bibfield
  {title} {\enquote {\bibinfo {title} {{Dissipation and forces in
  time-dependent Hartree-Fock calculations}},}\ }\href {\doibase
  10.1103/PhysRevC.37.1026} {\bibfield  {journal} {\bibinfo  {journal} {Phys.
  Rev. C}\ }\textbf {\bibinfo {volume} {37}},\ \bibinfo {pages} {1026}
  (\bibinfo {year} {1988})}\BibitemShut {NoStop}%
\bibitem [{\citenamefont {Stevenson}\ \emph {et~al.}(2016)\citenamefont
  {Stevenson}, \citenamefont {Suckling}, \citenamefont {Fracasso},
  \citenamefont {Barton},\ and\ \citenamefont {Umar}}]{Stevenson:2015dva}%
  \BibitemOpen
  \bibfield  {author} {\bibinfo {author} {\bibfnamefont {P.~D.}\ \bibnamefont
  {Stevenson}}, \bibinfo {author} {\bibfnamefont {E.~B.}\ \bibnamefont
  {Suckling}}, \bibinfo {author} {\bibfnamefont {S.}~\bibnamefont {Fracasso}},
  \bibinfo {author} {\bibfnamefont {M.~C.}\ \bibnamefont {Barton}}, \ and\
  \bibinfo {author} {\bibfnamefont {A.~S.}\ \bibnamefont {Umar}},\ }\bibfield
  {title} {\enquote {\bibinfo {title} {{Skyrme tensor force in heavy ion
  collisions}},}\ }\href {\doibase 10.1103/PhysRevC.93.054617} {\bibfield
  {journal} {\bibinfo  {journal} {Phys. Rev. C}\ }\textbf {\bibinfo {volume}
  {93}},\ \bibinfo {pages} {054617} (\bibinfo {year} {2016})}\BibitemShut
  {NoStop}%
\bibitem [{\citenamefont {Godbey}\ \emph {et~al.}(2019)\citenamefont {Godbey},
  \citenamefont {Guo},\ and\ \citenamefont {Umar}}]{Godbey:2019vlg}%
  \BibitemOpen
  \bibfield  {author} {\bibinfo {author} {\bibfnamefont {K.}~\bibnamefont
  {Godbey}}, \bibinfo {author} {\bibfnamefont {L.}~\bibnamefont {Guo}}, \ and\
  \bibinfo {author} {\bibfnamefont {A.~S.}\ \bibnamefont {Umar}},\ }\bibfield
  {title} {\enquote {\bibinfo {title} {{Influence of the tensor interaction on
  heavy-ion fusion cross sections}},}\ }\href {\doibase
  10.1103/PhysRevC.100.054612} {\bibfield  {journal} {\bibinfo  {journal}
  {Phys. Rev. C}\ }\textbf {\bibinfo {volume} {100}},\ \bibinfo {pages}
  {054612} (\bibinfo {year} {2019})}\BibitemShut {NoStop}%
\bibitem [{\citenamefont {Xu}\ and\ \citenamefont {Li}(2013)}]{Xu:2012hh}%
  \BibitemOpen
  \bibfield  {author} {\bibinfo {author} {\bibfnamefont {J.}~\bibnamefont
  {Xu}}\ and\ \bibinfo {author} {\bibfnamefont {B.~A.}\ \bibnamefont {Li}},\
  }\bibfield  {title} {\enquote {\bibinfo {title} {{Probing in-medium
  spin-orbit potential with intermediate-energy heavy-ion collisions}},}\
  }\href {\doibase 10.1016/j.physletb.2013.06.033} {\bibfield  {journal}
  {\bibinfo  {journal} {Phys. Lett. B}\ }\textbf {\bibinfo {volume} {724}},\
  \bibinfo {pages} {346} (\bibinfo {year} {2013})}\BibitemShut {NoStop}%
\bibitem [{\citenamefont {Xia}\ \emph {et~al.}(2016{\natexlab{a}})\citenamefont
  {Xia}, \citenamefont {Xu}, \citenamefont {Li},\ and\ \citenamefont
  {Shen}}]{Xia:2016xiw}%
  \BibitemOpen
  \bibfield  {author} {\bibinfo {author} {\bibfnamefont {Y.}~\bibnamefont
  {Xia}}, \bibinfo {author} {\bibfnamefont {J.}~\bibnamefont {Xu}}, \bibinfo
  {author} {\bibfnamefont {B.~A.}\ \bibnamefont {Li}}, \ and\ \bibinfo {author}
  {\bibfnamefont {W.~Q.}\ \bibnamefont {Shen}},\ }\bibfield  {title} {\enquote
  {\bibinfo {title} {{Equations of motion of test particles for solving the
  spin-dependent Boltzmann\textendash{}Vlasov equation}},}\ }\href {\doibase
  10.1016/j.physletb.2016.06.029} {\bibfield  {journal} {\bibinfo  {journal}
  {Phys. Lett. B}\ }\textbf {\bibinfo {volume} {759}},\ \bibinfo {pages} {596}
  (\bibinfo {year} {2016}{\natexlab{a}})}\BibitemShut {NoStop}%
\bibitem [{\citenamefont {Xia}\ \emph {et~al.}(2014)\citenamefont {Xia},
  \citenamefont {Xu}, \citenamefont {Li},\ and\ \citenamefont
  {Shen}}]{Xia:2014qva}%
  \BibitemOpen
  \bibfield  {author} {\bibinfo {author} {\bibfnamefont {Y.}~\bibnamefont
  {Xia}}, \bibinfo {author} {\bibfnamefont {J.}~\bibnamefont {Xu}}, \bibinfo
  {author} {\bibfnamefont {B.~A.}\ \bibnamefont {Li}}, \ and\ \bibinfo {author}
  {\bibfnamefont {W.~Q.}\ \bibnamefont {Shen}},\ }\bibfield  {title} {\enquote
  {\bibinfo {title} {{Spin-orbit coupling and the up-down differential
  transverse flow in intermediate-energy heavy-ion collisions}},}\ }\href
  {\doibase 10.1103/PhysRevC.89.064606} {\bibfield  {journal} {\bibinfo
  {journal} {Phys. Rev. C}\ }\textbf {\bibinfo {volume} {89}},\ \bibinfo
  {pages} {064606} (\bibinfo {year} {2014})}\BibitemShut {NoStop}%
\bibitem [{\citenamefont {Xia}\ and\ \citenamefont {Xu}(2020)}]{Xia:2019whr}%
  \BibitemOpen
  \bibfield  {author} {\bibinfo {author} {\bibfnamefont {Y.}~\bibnamefont
  {Xia}}\ and\ \bibinfo {author} {\bibfnamefont {J.}~\bibnamefont {Xu}},\
  }\bibfield  {title} {\enquote {\bibinfo {title} {{Nucleon spin polarization
  in intermediate-energy heavy-ion collisions}},}\ }\href {\doibase
  10.1016/j.physletb.2019.135130} {\bibfield  {journal} {\bibinfo  {journal}
  {Phys. Lett. B}\ }\textbf {\bibinfo {volume} {800}},\ \bibinfo {pages}
  {135130} (\bibinfo {year} {2020})}\BibitemShut {NoStop}%
\bibitem [{\citenamefont {Hirsch}(1999)}]{PhysRevLett.83.1834}%
  \BibitemOpen
  \bibfield  {author} {\bibinfo {author} {\bibfnamefont {J.~E.}\ \bibnamefont
  {Hirsch}},\ }\bibfield  {title} {\enquote {\bibinfo {title} {Spin hall
  effect},}\ }\href {\doibase 10.1103/PhysRevLett.83.1834} {\bibfield
  {journal} {\bibinfo  {journal} {Phys. Rev. Lett.}\ }\textbf {\bibinfo
  {volume} {83}},\ \bibinfo {pages} {1834} (\bibinfo {year}
  {1999})}\BibitemShut {NoStop}%
\bibitem [{\citenamefont {Ring}\ and\ \citenamefont {Schuck}(1980)}]{Ring1980}%
  \BibitemOpen
  \bibfield  {author} {\bibinfo {author} {\bibfnamefont {P.}~\bibnamefont
  {Ring}}\ and\ \bibinfo {author} {\bibfnamefont {P.}~\bibnamefont {Schuck}},\
  }\href@noop {} {\emph {\bibinfo {title} {The Nuclear Many-Body Problem}}}\
  (\bibinfo  {publisher} {Springer},\ \bibinfo {address} {Berlin},\ \bibinfo
  {year} {1980})\BibitemShut {NoStop}%
\bibitem [{\citenamefont {Smith}\ and\ \citenamefont
  {Jensen}(1989)}]{Smith1989}%
  \BibitemOpen
  \bibfield  {author} {\bibinfo {author} {\bibfnamefont {H.}~\bibnamefont
  {Smith}}\ and\ \bibinfo {author} {\bibfnamefont {H.H.}\ \bibnamefont
  {Jensen}},\ }\href@noop {} {\emph {\bibinfo {title} {Transport Phenomena}}}\
  (\bibinfo  {publisher} {Oxford University Press},\ \bibinfo {address}
  {Oxford},\ \bibinfo {year} {1989})\BibitemShut {NoStop}%
\bibitem [{\citenamefont {Vautherin}\ and\ \citenamefont
  {Brink}(1972)}]{Vautherin:1971aw}%
  \BibitemOpen
  \bibfield  {author} {\bibinfo {author} {\bibfnamefont {D.}~\bibnamefont
  {Vautherin}}\ and\ \bibinfo {author} {\bibfnamefont {D.~M.}\ \bibnamefont
  {Brink}},\ }\bibfield  {title} {\enquote {\bibinfo {title} {{Hartree-Fock
  calculations with Skyrme's interaction. 1. Spherical nuclei}},}\ }\href
  {\doibase 10.1103/PhysRevC.5.626} {\bibfield  {journal} {\bibinfo  {journal}
  {Phys. Rev. C}\ }\textbf {\bibinfo {volume} {5}},\ \bibinfo {pages} {626}
  (\bibinfo {year} {1972})}\BibitemShut {NoStop}%
\bibitem [{\citenamefont {Chen}\ \emph {et~al.}(2010)\citenamefont {Chen},
  \citenamefont {Ko}, \citenamefont {Li},\ and\ \citenamefont
  {Xu}}]{PhysRevC.82.024321}%
  \BibitemOpen
  \bibfield  {author} {\bibinfo {author} {\bibfnamefont {L.~W.}\ \bibnamefont
  {Chen}}, \bibinfo {author} {\bibfnamefont {C.~M.}\ \bibnamefont {Ko}},
  \bibinfo {author} {\bibfnamefont {B.~A.}\ \bibnamefont {Li}}, \ and\ \bibinfo
  {author} {\bibfnamefont {J.}~\bibnamefont {Xu}},\ }\bibfield  {title}
  {\enquote {\bibinfo {title} {Density slope of the nuclear symmetry energy
  from the neutron skin thickness of heavy nuclei},}\ }\href {\doibase
  10.1103/PhysRevC.82.024321} {\bibfield  {journal} {\bibinfo  {journal} {Phys.
  Rev. C}\ }\textbf {\bibinfo {volume} {82}},\ \bibinfo {pages} {024321}
  (\bibinfo {year} {2010})}\BibitemShut {NoStop}%
\bibitem [{\citenamefont {Engel}\ \emph {et~al.}(1975)\citenamefont {Engel},
  \citenamefont {Brink}, \citenamefont {Goeke}, \citenamefont {Krieger},\ and\
  \citenamefont {Vautherin}}]{Engel:1975zz}%
  \BibitemOpen
  \bibfield  {author} {\bibinfo {author} {\bibfnamefont {Y.~M.}\ \bibnamefont
  {Engel}}, \bibinfo {author} {\bibfnamefont {D.~M.}\ \bibnamefont {Brink}},
  \bibinfo {author} {\bibfnamefont {K.}~\bibnamefont {Goeke}}, \bibinfo
  {author} {\bibfnamefont {S.~J.}\ \bibnamefont {Krieger}}, \ and\ \bibinfo
  {author} {\bibfnamefont {D.}~\bibnamefont {Vautherin}},\ }\bibfield  {title}
  {\enquote {\bibinfo {title} {{Time-dependent hartree-fock theory with
  Skyrme's interaction}},}\ }\href {\doibase 10.1016/0375-9474(75)90184-0}
  {\bibfield  {journal} {\bibinfo  {journal} {Nucl. Phys. A}\ }\textbf
  {\bibinfo {volume} {249}},\ \bibinfo {pages} {215} (\bibinfo {year}
  {1975})}\BibitemShut {NoStop}%
\bibitem [{\citenamefont {Lenk}\ and\ \citenamefont
  {Pandharipande}(1989)}]{Lenk:1989zz}%
  \BibitemOpen
  \bibfield  {author} {\bibinfo {author} {\bibfnamefont {R.~J.}\ \bibnamefont
  {Lenk}}\ and\ \bibinfo {author} {\bibfnamefont {V.~R.}\ \bibnamefont
  {Pandharipande}},\ }\bibfield  {title} {\enquote {\bibinfo {title} {{Nuclear
  mean field dynamics in the lattice Hamiltonian Vlasov method}},}\ }\href
  {\doibase 10.1103/PhysRevC.39.2242} {\bibfield  {journal} {\bibinfo
  {journal} {Phys. Rev. C}\ }\textbf {\bibinfo {volume} {39}},\ \bibinfo
  {pages} {2242} (\bibinfo {year} {1989})}\BibitemShut {NoStop}%
\bibitem [{\citenamefont {Xia}\ \emph {et~al.}(2017)\citenamefont {Xia},
  \citenamefont {Xu}, \citenamefont {Li},\ and\ \citenamefont
  {Shen}}]{Xia:2017dbx}%
  \BibitemOpen
  \bibfield  {author} {\bibinfo {author} {\bibfnamefont {Y.}~\bibnamefont
  {Xia}}, \bibinfo {author} {\bibfnamefont {J.}~\bibnamefont {Xu}}, \bibinfo
  {author} {\bibfnamefont {B.~A.}\ \bibnamefont {Li}}, \ and\ \bibinfo {author}
  {\bibfnamefont {W.~Q.}\ \bibnamefont {Shen}},\ }\bibfield  {title} {\enquote
  {\bibinfo {title} {{Simulating spin dynamics with spin-dependent cross
  sections in heavy-ion collisions}},}\ }\href {\doibase
  10.1103/PhysRevC.96.044618} {\bibfield  {journal} {\bibinfo  {journal} {Phys.
  Rev. C}\ }\textbf {\bibinfo {volume} {96}},\ \bibinfo {pages} {044618}
  (\bibinfo {year} {2017})}\BibitemShut {NoStop}%
\bibitem [{\citenamefont {Arndt}\ \emph {et~al.}(1977)\citenamefont {Arndt},
  \citenamefont {Hackman},\ and\ \citenamefont {Roper}}]{PhysRevC.15.1002}%
  \BibitemOpen
  \bibfield  {author} {\bibinfo {author} {\bibfnamefont {R.~A.}\ \bibnamefont
  {Arndt}}, \bibinfo {author} {\bibfnamefont {R.~H.}\ \bibnamefont {Hackman}},
  \ and\ \bibinfo {author} {\bibfnamefont {L.~D.}\ \bibnamefont {Roper}},\
  }\bibfield  {title} {\enquote {\bibinfo {title} {Nucleon-nucleon scattering
  analyses. ii. neutron-proton scattering from 0 to 425 mev and proton-proton
  scattering from 1 to 500 mev},}\ }\href {\doibase 10.1103/PhysRevC.15.1002}
  {\bibfield  {journal} {\bibinfo  {journal} {Phys. Rev. C}\ }\textbf {\bibinfo
  {volume} {15}},\ \bibinfo {pages} {1002} (\bibinfo {year}
  {1977})}\BibitemShut {NoStop}%
\bibitem [{\citenamefont {Bell}\ and\ \citenamefont
  {Skyrme}(1956)}]{Bell1956-BELCTN}%
  \BibitemOpen
  \bibfield  {author} {\bibinfo {author} {\bibfnamefont {J.~S.}\ \bibnamefont
  {Bell}}\ and\ \bibinfo {author} {\bibfnamefont {T.~H.~R.}\ \bibnamefont
  {Skyrme}},\ }\bibfield  {title} {\enquote {\bibinfo {title} {Cviii. the
  nuclear spin-orbit coupling},}\ }\href {\doibase 10.1080/14786435608238187}
  {\bibfield  {journal} {\bibinfo  {journal} {Philosophical Magazine}\ }\textbf
  {\bibinfo {volume} {1}},\ \bibinfo {pages} {1055} (\bibinfo {year}
  {1956})}\BibitemShut {NoStop}%
\bibitem [{\citenamefont {Wiringa}\ \emph {et~al.}(1995)\citenamefont
  {Wiringa}, \citenamefont {Stoks},\ and\ \citenamefont
  {Schiavilla}}]{Wiringa:1994wb}%
  \BibitemOpen
  \bibfield  {author} {\bibinfo {author} {\bibfnamefont {R.~B.}\ \bibnamefont
  {Wiringa}}, \bibinfo {author} {\bibfnamefont {V.~G.~J.}\ \bibnamefont
  {Stoks}}, \ and\ \bibinfo {author} {\bibfnamefont {R.}~\bibnamefont
  {Schiavilla}},\ }\bibfield  {title} {\enquote {\bibinfo {title} {{An Accurate
  nucleon-nucleon potential with charge independence breaking}},}\ }\href
  {\doibase 10.1103/PhysRevC.51.38} {\bibfield  {journal} {\bibinfo  {journal}
  {Phys. Rev. C}\ }\textbf {\bibinfo {volume} {51}},\ \bibinfo {pages} {38}
  (\bibinfo {year} {1995})}\BibitemShut {NoStop}%
\bibitem [{\citenamefont {Sammarruca}\ \emph {et~al.}(1999)\citenamefont
  {Sammarruca}, \citenamefont {Stephenson}, \citenamefont {Jiang},
  \citenamefont {Liu}, \citenamefont {Olmer}, \citenamefont {Opper},\ and\
  \citenamefont {Wissink}}]{PhysRevC.61.014309}%
  \BibitemOpen
  \bibfield  {author} {\bibinfo {author} {\bibfnamefont {F.}~\bibnamefont
  {Sammarruca}}, \bibinfo {author} {\bibfnamefont {E.~J.}\ \bibnamefont
  {Stephenson}}, \bibinfo {author} {\bibfnamefont {K.}~\bibnamefont {Jiang}},
  \bibinfo {author} {\bibfnamefont {J.}~\bibnamefont {Liu}}, \bibinfo {author}
  {\bibfnamefont {C.}~\bibnamefont {Olmer}}, \bibinfo {author} {\bibfnamefont
  {A.~K.}\ \bibnamefont {Opper}}, \ and\ \bibinfo {author} {\bibfnamefont
  {S.~W.}\ \bibnamefont {Wissink}},\ }\bibfield  {title} {\enquote {\bibinfo
  {title} {Testing microscopic medium effects on nucleons and mesons using
  polarization observables in high-spin, unnatural-parity
  $(\stackrel{\ensuremath{\rightarrow}}{p},{p}^{\ensuremath{'}})$ reactions at
  200 mev},}\ }\href {\doibase 10.1103/PhysRevC.61.014309} {\bibfield
  {journal} {\bibinfo  {journal} {Phys. Rev. C}\ }\textbf {\bibinfo {volume}
  {61}},\ \bibinfo {pages} {014309} (\bibinfo {year} {1999})}\BibitemShut
  {NoStop}%
\bibitem [{\citenamefont {Gale}\ and\ \citenamefont
  {Das~Gupta}(1990)}]{Gale:1990zz}%
  \BibitemOpen
  \bibfield  {author} {\bibinfo {author} {\bibfnamefont {C.}~\bibnamefont
  {Gale}}\ and\ \bibinfo {author} {\bibfnamefont {S.}~\bibnamefont
  {Das~Gupta}},\ }\bibfield  {title} {\enquote {\bibinfo {title} {{Conservation
  laws and nuclear transport models}},}\ }\href {\doibase
  10.1103/PhysRevC.42.1577} {\bibfield  {journal} {\bibinfo  {journal} {Phys.
  Rev. C}\ }\textbf {\bibinfo {volume} {42}},\ \bibinfo {pages} {1577}
  (\bibinfo {year} {1990})}\BibitemShut {NoStop}%
\bibitem [{\citenamefont {Liu}\ and\ \citenamefont {Xu}(2023)}]{Liu:2023pgc}%
  \BibitemOpen
  \bibfield  {author} {\bibinfo {author} {\bibfnamefont {R.~J.}\ \bibnamefont
  {Liu}}\ and\ \bibinfo {author} {\bibfnamefont {J.}~\bibnamefont {Xu}},\
  }\bibfield  {title} {\enquote {\bibinfo {title} {{Revisiting angular momentum
  conservation in transport simulations of intermediate-energy heavy-ion
  collisions}},}\ }\href {\doibase 10.3390/universe9010036} {\bibfield
  {journal} {\bibinfo  {journal} {Universe}\ }\textbf {\bibinfo {volume} {9}},\
  \bibinfo {pages} {36} (\bibinfo {year} {2023})}\BibitemShut {NoStop}%
\bibitem [{\citenamefont {Papa}\ \emph {et~al.}(2005)\citenamefont {Papa},
  \citenamefont {Giuliani},\ and\ \citenamefont {Bonasera}}]{Papa:2005sp}%
  \BibitemOpen
  \bibfield  {author} {\bibinfo {author} {\bibfnamefont {M.}~\bibnamefont
  {Papa}}, \bibinfo {author} {\bibfnamefont {G.}~\bibnamefont {Giuliani}}, \
  and\ \bibinfo {author} {\bibfnamefont {A.}~\bibnamefont {Bonasera}},\
  }\bibfield  {title} {\enquote {\bibinfo {title} {{Constrained molecular
  dynamics II: A N-body approach to nuclear systems}},}\ }\href {\doibase
  10.1016/j.jcp.2005.02.032} {\bibfield  {journal} {\bibinfo  {journal} {J.
  Comput. Phys.}\ }\textbf {\bibinfo {volume} {208}},\ \bibinfo {pages} {403}
  (\bibinfo {year} {2005})}\BibitemShut {NoStop}%
\bibitem [{\citenamefont {Xia}\ \emph {et~al.}(2016{\natexlab{b}})\citenamefont
  {Xia}, \citenamefont {Xu}, \citenamefont {Li},\ and\ \citenamefont
  {Shen}}]{Xia:2014rua}%
  \BibitemOpen
  \bibfield  {author} {\bibinfo {author} {\bibfnamefont {Y.}~\bibnamefont
  {Xia}}, \bibinfo {author} {\bibfnamefont {J.}~\bibnamefont {Xu}}, \bibinfo
  {author} {\bibfnamefont {B.~A.}\ \bibnamefont {Li}}, \ and\ \bibinfo {author}
  {\bibfnamefont {W.~Q.}\ \bibnamefont {Shen}},\ }\bibfield  {title} {\enquote
  {\bibinfo {title} {{Spin transport in intermediate-energy heavy-ion
  collisions as a probe of in-medium spin\textendash{}orbit interactions}},}\
  }\href {\doibase 10.1016/j.nuclphysa.2016.06.001} {\bibfield  {journal}
  {\bibinfo  {journal} {Nucl. Phys. A}\ }\textbf {\bibinfo {volume} {955}},\
  \bibinfo {pages} {41} (\bibinfo {year} {2016}{\natexlab{b}})}\BibitemShut
  {NoStop}%
\bibitem [{\citenamefont {Chen}\ \emph {et~al.}(2003)\citenamefont {Chen},
  \citenamefont {Ko},\ and\ \citenamefont {Li}}]{Chen:2003ava}%
  \BibitemOpen
  \bibfield  {author} {\bibinfo {author} {\bibfnamefont {L.~W.}\ \bibnamefont
  {Chen}}, \bibinfo {author} {\bibfnamefont {C.~M.}\ \bibnamefont {Ko}}, \ and\
  \bibinfo {author} {\bibfnamefont {B.~A.}\ \bibnamefont {Li}},\ }\bibfield
  {title} {\enquote {\bibinfo {title} {{Light cluster production in
  intermediate-energy heavy ion collisions induced by neutron rich nuclei}},}\
  }\href {\doibase 10.1016/j.nuclphysa.2003.09.010} {\bibfield  {journal}
  {\bibinfo  {journal} {Nucl. Phys. A}\ }\textbf {\bibinfo {volume} {729}},\
  \bibinfo {pages} {809} (\bibinfo {year} {2003})}\BibitemShut {NoStop}%
\bibitem [{\citenamefont {Sun}\ and\ \citenamefont {Chen}(2017)}]{Sun:2017ooe}%
  \BibitemOpen
  \bibfield  {author} {\bibinfo {author} {\bibfnamefont {K.~J.}\ \bibnamefont
  {Sun}}\ and\ \bibinfo {author} {\bibfnamefont {L.~W.}\ \bibnamefont {Chen}},\
  }\bibfield  {title} {\enquote {\bibinfo {title} {{Analytical coalescence
  formula for particle production in relativistic heavy-ion collisions}},}\
  }\href {\doibase 10.1103/PhysRevC.95.044905} {\bibfield  {journal} {\bibinfo
  {journal} {Phys. Rev. C}\ }\textbf {\bibinfo {volume} {95}},\ \bibinfo
  {pages} {044905} (\bibinfo {year} {2017})}\BibitemShut {NoStop}%
\bibitem [{\citenamefont {Ropke}(2009)}]{Ropke:2008qk}%
  \BibitemOpen
  \bibfield  {author} {\bibinfo {author} {\bibfnamefont {G.}~\bibnamefont
  {Ropke}},\ }\bibfield  {title} {\enquote {\bibinfo {title} {{Light nuclei
  quasiparticle energy shift in hot and dense nuclear matter}},}\ }\href
  {\doibase 10.1103/PhysRevC.79.014002} {\bibfield  {journal} {\bibinfo
  {journal} {Phys. Rev. C}\ }\textbf {\bibinfo {volume} {79}},\ \bibinfo
  {pages} {014002} (\bibinfo {year} {2009})}\BibitemShut {NoStop}%
\bibitem [{\citenamefont {Liang}\ and\ \citenamefont
  {Wang}(2005)}]{Liang:2004xn}%
  \BibitemOpen
  \bibfield  {author} {\bibinfo {author} {\bibfnamefont {Z.~T.}\ \bibnamefont
  {Liang}}\ and\ \bibinfo {author} {\bibfnamefont {X.~N.}\ \bibnamefont
  {Wang}},\ }\bibfield  {title} {\enquote {\bibinfo {title} {{Spin alignment of
  vector mesons in non-central A+A collisions}},}\ }\href {\doibase
  10.1016/j.physletb.2005.09.060} {\bibfield  {journal} {\bibinfo  {journal}
  {Phys. Lett. B}\ }\textbf {\bibinfo {volume} {629}},\ \bibinfo {pages} {20}
  (\bibinfo {year} {2005})}\BibitemShut {NoStop}%
\bibitem [{\citenamefont {Adamczyk}\ \emph {et~al.}(2017)\citenamefont
  {Adamczyk} \emph {et~al.}}]{STAR:2017ckg}%
  \BibitemOpen
  \bibfield  {author} {\bibinfo {author} {\bibfnamefont {L.}~\bibnamefont
  {Adamczyk}} \emph {et~al.} (\bibinfo {collaboration} {STAR}),\ }\bibfield
  {title} {\enquote {\bibinfo {title} {{Global $\Lambda$ hyperon polarization
  in nuclear collisions: evidence for the most vortical fluid}},}\ }\href
  {\doibase 10.1038/nature23004} {\bibfield  {journal} {\bibinfo  {journal}
  {Nature}\ }\textbf {\bibinfo {volume} {548}},\ \bibinfo {pages} {62}
  (\bibinfo {year} {2017})}\BibitemShut {NoStop}%
\bibitem [{\citenamefont {Adam~{\it et al.}}(2019)}]{PhysRevLett.123.132301}%
  \BibitemOpen
  \bibfield  {author} {\bibinfo {author} {\bibfnamefont {J.}~\bibnamefont
  {Adam~{\it et al.}}} (\bibinfo {collaboration} {STAR Collaboration}),\
  }\bibfield  {title} {\enquote {\bibinfo {title} {Polarization of
  $\mathrm{\ensuremath{\Lambda}}$ ($\overline{\mathrm{\ensuremath{\Lambda}}}$)
  hyperons along the beam direction in $\mathrm{Au}+\mathrm{Au}$ collisions at
  $\sqrt{{s}_{NN}}=200\text{ }\text{ }\mathrm{GeV}$},}\ }\href {\doibase
  10.1103/PhysRevLett.123.132301} {\bibfield  {journal} {\bibinfo  {journal}
  {Phys. Rev. Lett.}\ }\textbf {\bibinfo {volume} {123}},\ \bibinfo {pages}
  {132301} (\bibinfo {year} {2019})}\BibitemShut {NoStop}%
\bibitem [{\citenamefont {Becattini}\ \emph {et~al.}(2021)\citenamefont
  {Becattini}, \citenamefont {Buzzegoli},\ and\ \citenamefont
  {Palermo}}]{Becattini:2021suc}%
  \BibitemOpen
  \bibfield  {author} {\bibinfo {author} {\bibfnamefont {F.}~\bibnamefont
  {Becattini}}, \bibinfo {author} {\bibfnamefont {M.}~\bibnamefont
  {Buzzegoli}}, \ and\ \bibinfo {author} {\bibfnamefont {A.}~\bibnamefont
  {Palermo}},\ }\bibfield  {title} {\enquote {\bibinfo {title} {{Spin-thermal
  shear coupling in a relativistic fluid}},}\ }\href {\doibase
  10.1016/j.physletb.2021.136519} {\bibfield  {journal} {\bibinfo  {journal}
  {Phys. Lett. B}\ }\textbf {\bibinfo {volume} {820}},\ \bibinfo {pages}
  {136519} (\bibinfo {year} {2021})}\BibitemShut {NoStop}%
\bibitem [{\citenamefont {Fu}\ \emph {et~al.}(2021)\citenamefont {Fu},
  \citenamefont {Liu}, \citenamefont {Pang}, \citenamefont {Song},\ and\
  \citenamefont {Yin}}]{Fu:2021pok}%
  \BibitemOpen
  \bibfield  {author} {\bibinfo {author} {\bibfnamefont {B.}~\bibnamefont
  {Fu}}, \bibinfo {author} {\bibfnamefont {S.~Y.~F.}\ \bibnamefont {Liu}},
  \bibinfo {author} {\bibfnamefont {L.}~\bibnamefont {Pang}}, \bibinfo {author}
  {\bibfnamefont {H.}~\bibnamefont {Song}}, \ and\ \bibinfo {author}
  {\bibfnamefont {Y.}~\bibnamefont {Yin}},\ }\bibfield  {title} {\enquote
  {\bibinfo {title} {{Shear-Induced Spin Polarization in Heavy-Ion
  Collisions}},}\ }\href {\doibase 10.1103/PhysRevLett.127.142301} {\bibfield
  {journal} {\bibinfo  {journal} {Phys. Rev. Lett.}\ }\textbf {\bibinfo
  {volume} {127}},\ \bibinfo {pages} {142301} (\bibinfo {year}
  {2021})}\BibitemShut {NoStop}%
\bibitem [{\citenamefont {Becattini}\ and\ \citenamefont
  {Karpenko}(2018)}]{PhysRevLett.120.012302}%
  \BibitemOpen
  \bibfield  {author} {\bibinfo {author} {\bibfnamefont {F.}~\bibnamefont
  {Becattini}}\ and\ \bibinfo {author} {\bibfnamefont {Iu.}\ \bibnamefont
  {Karpenko}},\ }\bibfield  {title} {\enquote {\bibinfo {title} {Collective
  longitudinal polarization in relativistic heavy-ion collisions at very high
  energy},}\ }\href {\doibase 10.1103/PhysRevLett.120.012302} {\bibfield
  {journal} {\bibinfo  {journal} {Phys. Rev. Lett.}\ }\textbf {\bibinfo
  {volume} {120}},\ \bibinfo {pages} {012302} (\bibinfo {year}
  {2018})}\BibitemShut {NoStop}%
\bibitem [{\citenamefont {Liu}\ \emph {et~al.}(2020)\citenamefont {Liu},
  \citenamefont {Sun},\ and\ \citenamefont {Ko}}]{PhysRevLett.125.062301}%
  \BibitemOpen
  \bibfield  {author} {\bibinfo {author} {\bibfnamefont {S.~Y.~F.}\
  \bibnamefont {Liu}}, \bibinfo {author} {\bibfnamefont {Y.}~\bibnamefont
  {Sun}}, \ and\ \bibinfo {author} {\bibfnamefont {C.~M.}\ \bibnamefont {Ko}},\
  }\bibfield  {title} {\enquote {\bibinfo {title} {Spin polarizations in a
  covariant angular-momentum-conserved chiral transport model},}\ }\href
  {\doibase 10.1103/PhysRevLett.125.062301} {\bibfield  {journal} {\bibinfo
  {journal} {Phys. Rev. Lett.}\ }\textbf {\bibinfo {volume} {125}},\ \bibinfo
  {pages} {062301} (\bibinfo {year} {2020})}\BibitemShut {NoStop}%
\bibitem [{\citenamefont {Sheng}\ \emph {et~al.}(2020)\citenamefont {Sheng},
  \citenamefont {Oliva},\ and\ \citenamefont {Wang}}]{Sheng:2019kmk}%
  \BibitemOpen
  \bibfield  {author} {\bibinfo {author} {\bibfnamefont {X.~L.}\ \bibnamefont
  {Sheng}}, \bibinfo {author} {\bibfnamefont {L.}~\bibnamefont {Oliva}}, \ and\
  \bibinfo {author} {\bibfnamefont {Q.}~\bibnamefont {Wang}},\ }\bibfield
  {title} {\enquote {\bibinfo {title} {{What can we learn from the global spin
  alignment of $\phi$ mesons in heavy-ion collisions?}}}\ }\href {\doibase
  10.1103/PhysRevD.101.096005} {\bibfield  {journal} {\bibinfo  {journal}
  {Phys. Rev. D}\ }\textbf {\bibinfo {volume} {101}},\ \bibinfo {pages}
  {096005} (\bibinfo {year} {2020})},\ \bibinfo {note} {[Erratum: Phys.Rev.D
  105, 099903 (2022)]}\BibitemShut {NoStop}%
\bibitem [{\citenamefont {Sheng}\ \emph {et~al.}(2023)\citenamefont {Sheng},
  \citenamefont {Oliva}, \citenamefont {Liang}, \citenamefont {Wang},\ and\
  \citenamefont {Wang}}]{Sheng:2022wsy}%
  \BibitemOpen
  \bibfield  {author} {\bibinfo {author} {\bibfnamefont {X.~L.}\ \bibnamefont
  {Sheng}}, \bibinfo {author} {\bibfnamefont {L.}~\bibnamefont {Oliva}},
  \bibinfo {author} {\bibfnamefont {Z.~T.}\ \bibnamefont {Liang}}, \bibinfo
  {author} {\bibfnamefont {Q.}~\bibnamefont {Wang}}, \ and\ \bibinfo {author}
  {\bibfnamefont {X.~N.}\ \bibnamefont {Wang}},\ }\bibfield  {title} {\enquote
  {\bibinfo {title} {{Spin Alignment of Vector Mesons in Heavy-Ion
  Collisions}},}\ }\href {\doibase 10.1103/PhysRevLett.131.042304} {\bibfield
  {journal} {\bibinfo  {journal} {Phys. Rev. Lett.}\ }\textbf {\bibinfo
  {volume} {131}},\ \bibinfo {pages} {042304} (\bibinfo {year}
  {2023})}\BibitemShut {NoStop}%
\end{thebibliography}%
\end{document}